\documentclass[%
 reprint,
 prd,
superscriptaddress,
 amsmath,amssymb,
 aps,
  nofootinbib,
]{revtex4-1}

\usepackage{comment}
\usepackage{dsfont}
\usepackage{graphicx}
\usepackage{dcolumn}
\usepackage{bm}
\usepackage{color}
\usepackage{epsfig}
\usepackage[colorlinks=true, linkcolor=blue, citecolor=blue, urlcolor=blue]{hyperref}
\usepackage{natbib}
\usepackage{float}
\usepackage{footmisc}
\usepackage{slashed}
\usepackage[dvipsnames]{xcolor}
\usepackage{aas_macros}
\definecolor{lightred}{rgb}{1,0.5,0.5}
\definecolor{lightgreen}{rgb}{0.5,1,0.5}
\definecolor{lightblue}{rgb}{0.5,0.5,1}
\definecolor{lightcyan}{rgb}{0.5,0.75,0.75}
\definecolor{lightmagenta}{rgb}{0.75,0.5,0.75}
\definecolor{customgreen}{rgb}{0.494,1,0.502}

\usepackage{soul}

\newcommand{\meV}{\mathinner{\mathrm{meV}}}
\newcommand{\eV}{\mathinner{\mathrm{eV}}}
\newcommand{\keV}{\mathinner{\mathrm{keV}}}
\newcommand{\MeV}{\mathinner{\mathrm{MeV}}}
\newcommand{\GeV}{\mathinner{\mathrm{GeV}}}

\begin{document}
\title{Making the Subdominant Dominant: \\Gravothermal Pile-Up of Collisional Dark Matter Around Compact Objects}

\author{Reza Ebadi}
\email{ebadi@umd.edu}
\affiliation{Department of Physics, University of Maryland, College Park, Maryland 20742, USA}
\author{Erwin H.~Tanin}
\email{ehtanin@stanford.edu}
\affiliation{Stanford Institute for Theoretical Physics, Stanford University, Stanford, California 94305, USA}
\begin{abstract}
    The dark matter may consist of multiple species that interact differently. We show that a species that is cosmologically subdominant but highly self-collisional can pile up and become dominant in deep gravitational wells, such as those of white dwarfs and neutron stars.
\end{abstract}

\maketitle

\section{Introduction}
Proposed extensions to the Standard Model (SM) often include new particles, either as necessary ingredients or unintended byproducts. Some of these particles are expected to account for the observed dark matter (DM) abundance in the universe. Their self-interaction cross-section-to-mass ratios are constrained to be
$\sigma_m\lesssim 1\text{ cm}^2/\text{g}$ \cite{Peter:2012jh, Randall:2008ppe, Markevitch:2003at}, with a possible preference toward the upper end if the self-interacting DM (SIDM) picture \cite{Spergel:1999mh,Tulin:2017ara,Kaplinghat:2019dhn} holds \cite{Bullock:2017xww,2019MNRAS.487.1380G}. The rest generically constitute tiny ($\lesssim 10\%$) subcomponents of the total DM and have virtually unconstrained self-interactions. In this paper, we demonstrate that subcomponents that are cosmologically insignificant can rise to importance locally, e.g., around compact objects, if they self-interact sufficiently strongly.

We consider a subcomponent with arguably the simplest type of self-interaction, namely elastic and velocity-independent collision, but with an extremely large $\sigma_m$. Motivated by $\sigma_m$ values found in the SM, we focus on $\sigma_m$ in the range $10$-$10^{10}\text{ cm}^2/\text{g}$, although larger values are not ruled out observationally. The lower and upper ends of this range are close to the $\sigma_m$ of a nucleus and a molecule. We will show that subcomponents with such a large $\sigma_m$ pile up in deep gravitational wells, so much so that they can far dominate the DM mass density locally. Thermal pressure poses a hurdle to the pile-up, but heat conduction enabled by elastic collisions relieves the pressure support gradually, allowing the piling to continue. This mechanism, we dub \textit{gravothermal pile up}, may occur in many setups. Here, we focus on  white dwarfs (WDs) and neutron stars (NSs) in galactic environment.

Scenarios where subcomponents of DM self-interact appreciably have previously been explored. 
Models with large elastic $\sigma_m$, akin to the type we consider, have been proposed to explain the origin of supermassive black holes \cite{Pollack:2014rja,Roberts:2024wup,Choquette:2018lvq}, but they are yet to be explored at much smaller length scales. Other works focused mainly on subcomponents with dissipative self-interactions, a possibility largely motivated by the mirror-world scenario \cite{Mohapatra:2001sx,Foot:2014uba,Chacko:2005pe,Chacko:2005vw,Okun:2006eb,Berezhiani:2003xm,Foot:2014mia,Foot:2004pa,Ciarcelluti:2010zz,Foot:2016wvj} and often called partially interacting dark matter \cite{Fan:2013yva,Fan:2013tia}. Consequences of a self-interacting dark sector include the formation of a dark disk \cite{Fan:2013yva}, novel indirect-detection signatures \cite{Fan:2013yva,Agrawal:2017pnb}, dark acoustic oscillations \cite{Chacko:2016kgg,Cyr-Racine:2013fsa}, and formation of compact structures \cite{Buckley:2017ttd, Ghalsasi:2017jna, Chang:2018bgx}, among others \cite{Roy:2024bcu,Gemmell:2023trd,Roy:2023zar,Geller:2022gey,Chacko:2021vin,Mohapatra:1996yy,Yang:2025dgl,Yang:2025xsp}.

In the following, we adopt natural units: $c = k_B = 1$.

\begin{figure}[!t]
    \centering
    \includegraphics[width=1\columnwidth]{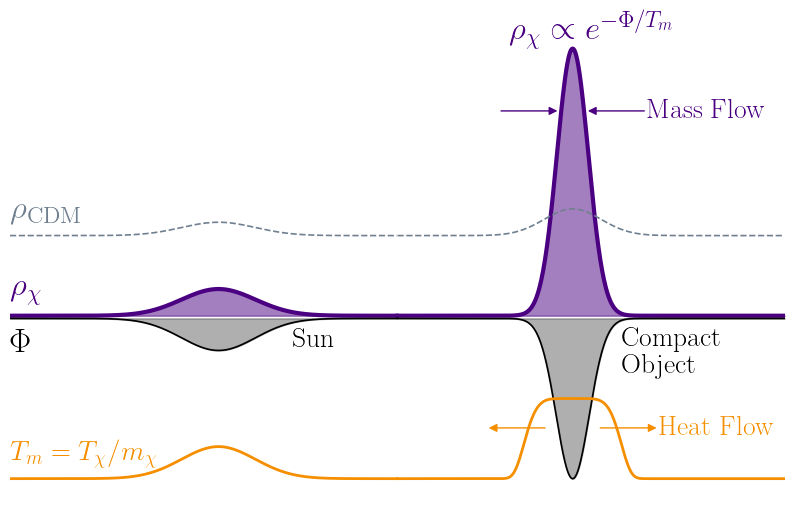}
    \caption{Gravothermal pile-up mechanism. Collisionless cold dark matter (CDM) transiting through a gravitational well gets mildly focused. Collisional DM subcomponent $\chi$ instead piles up gravothermally: as heat flows out through conduction, decreased pressure lets more $\chi$ particles to flow in. Around a sufficiently deep gravitational well $\Phi$, the subcomponent $\chi$ can be locally dominant over the CDM.}
    \label{fig:cartoon}
\end{figure}

\section{Boltzmann's Equilibrium}

Given an ambient gas of collisional particles $\chi$ with mass density $\rho_\chi^\infty$ and temperature per unit mass $T_m^\infty$, we introduce a fixed central gravitational well $\Phi(r)<0$ that vanishes at infinity. It is well known that the gas will seek the Boltzmann-enhanced density profile $\rho_\chi^{\rm eq}(r)=\rho_\chi^\infty e^{-\Phi(r)/T_m^\infty}$, which could be substantially enhanced if $-\Phi\gg T_m^\infty$, reflecting a propensity of the gas to collect in the well. This density profile is that of an ideal gas in hydrostatic equilibrium, if it could establish a global thermal equilibrium, $T_m(r)=T_m^\infty$. In reality, global thermalization takes time, and the gas may only thermalize locally, with some temperature profile $T_m(r)>T_m^\infty$. For each $T_m(r)$, there is a unique density profile that yields hydrostatic equilibrium,
\begin{align}
    \rho_\chi(r)=\rho_\chi^\infty\frac{T_m^\infty}{T_m(r)}e^{-\int_{\infty}^r dr'\, \frac{\partial_{r'}\Phi(r')}{T_m(r')}} \label{eq:hydrostatic}
\end{align}
If there exists a region $r\leq r_{\rm iso}$ where the temperature is approximately uniform, $T_m(r)\approx T_m^{\rm iso}> T_m^\infty$, the density profile inside this isothermal region is $\rho_\chi^{\rm iso}(r)\propto e^{-\Phi(r)/T_m^{\rm iso}}$, which is similar to $\rho_\chi^{\rm eq}$ apart from a prefactor that depends on profiles exterior to the isothermal region. Therefore, in this case an exponentially enhanced overdensity could still occur if $-\Phi/T_m^{\rm iso}\gtrsim \text{few}$.

To what degree the global thermal equilibrium is approached is a dynamical, history-dependent question. When the gas first settles into a hydrostatic equilibrium, its temperature typically rises to $T_m\sim -\Phi$, through a combination of shock and compressional heating. Subsequently, the system evolves toward global thermalization gravothermally: collisions cause heat to flow outward through conduction, cooling the central gas gradually and allowing mass to flow inward to re-establish hydrostatic equilibrium with higher density, as per Eq.~\eqref{eq:hydrostatic}. This gravothermal pile-up process is illustrated in Fig.~\ref{fig:cartoon}. The final temperature profile and, correspondingly, the amount of density enhancement will be determined by the efficiency of the heat conduction.

Significant gravothermal pile-up may occur around a potential whose escape velocity is $v_{\rm esc}$ if the environment has $T_m^\infty \ll v_{\rm esc}^2/2$. As a start, we will adopt a galactic environment, where typically $T_m^\infty\sim 10^{-6}$. For instance, in the Milky Way the Sun's core already satisfies the requisite condition with its $v_{\rm esc}^2/2\approx G(0.5M_\odot)/(0.2 R_\odot)=5\times 10^{-6}$, albeit marginally. Any solar-mass objects more compact than the Sun are even better sites for gravothermal pile-up. We will focus on WDs and NSs with benchmark properties shown in Table~\ref{tab:co_sources}.

\begin{table}[t!]
\small
\caption{Benchmark properties of compact objects.}\label{tab:co_sources}%
\begin{center}
\begin{tabular}{l|c|c|c|c}\hline\hline
\begin{tabular}{@{}c@{}}{}\\ \end{tabular} & 
\begin{tabular}{@{}c@{}}$M_{\star}\,[M_\odot]$ \\ \end{tabular} & 
\begin{tabular}{@{}c@{}}$R_{\star}\,[\text{km}]$ \\ \end{tabular} &
\begin{tabular}{@{}c@{}}$v_{\rm esc}$ \\ \end{tabular} &
\begin{tabular}{@{}c@{}}$t_{\rm age}\text{ [Gyr]}$ \\ \end{tabular}\\ \hline\hline
{} & {} & {} & {} \\
white dwarf &$1$ & $6000$ & 0.02& 10\\
neutron star &$1.4$ & $10$ &0.6 & 10\\
{} & {} & {} & {}\\
\hline
\end{tabular}
\end{center}
\end{table}

\section{Setup}
We assume that particles $\chi$ with cross-section-to-mass ratio $\sigma_m\gg 1\text{ cm}^2/\text{g}$ are present in galaxies with typical density and temperature per unit mass
\begin{align}
    \rho_\chi^\infty=f\times 0.4\GeV/\text{cm}^3,\quad T_m^\infty=10^{-6}
\end{align}
where $f\lesssim 10\%$ and $0.4\GeV/\text{cm}^3$ is the local DM density \cite{deSalas:2020hbh}. During structure formation, the dominant DM provides potential wells that the $\chi$ gas falls into. \textit{A priori}, due to its collisional nature, the $\chi$ distribution need not necessarily follow that of the dominant DM at the scales of nonlinear structures. This leads to the question of how the subcomponent fraction in a typical DM halo $f$ relates to the cosmic fraction $\bar{f}=\Omega_\chi/\Omega_{\rm DM}$. We expect the subcomponent to behave as non-radiative perfect fluid on DM-halo scales,\footnote{The mean free path of the subcomponent is negligible at the scale of $10\text{ kpc}$ if $f\sigma_m\gtrsim 100\text{ cm}^2/\text{g}$. This is satisfied in nearly all the parameter space we consider, except for the $f\sigma_m\sim 1-100\text{ cm}^2/\text{g}$ range. Since the efficiency of heat conduction peaks when the mean free path matches the system size, the subcomponent in that regime may behave very differently from perfect fluid. Nevertheless, this will not affect our results as we will express them in terms of the galactic fraction $f$ rather than the cosmic fraction $\bar{f}$.} much like the SM baryon gas, but with the radiative cooling turned off. Incidentally, simulations of structure formation in vanilla cosmology with a non-radiative fluid subcomponent added to the mix have been performed extensively in the past. They were used to test assumptions on the main mechanisms determining the baryon fraction of DM halos, including the roles of gas cooling. 
In short, these simulations suggest that $f \simeq 0.9\bar{f}$ across a wide range of DM halo masses; from $10^{15}\,M_\odot$ (clusters) to $10^{10}\,M_\odot$ (dwarfs) \cite{Crain:2006sb}, and extending down to $10^7\,M_\odot$ (ultra-faint dwarfs) and even $10^4\,M_\odot$ (minihalos) \cite{Zheng:2024xcc}. These studies also indicate that the subcomponent’s density distributions within halos follow the standard NFW profile.\footnote{The results of these simulations depend on the initial cosmic temperature of the subcomponent prior to structure formation; higher temperatures tend to suppress its fraction in the central region of the CDM halo. For initial temperatures comparable to or less than that of realistic baryonic gas ($\sim 10\meV$) the suppression is not significant \cite{2012JCAP...10..047A,Zheng:2024xcc}. Also, these simulations are not to be confused with those that studied the fluid limit of SIDM as the \textit{dominant} DM, which found that the resulting halos resemble cuspy isothermal spheres, with  density profiles $\rho_\chi(r)\propto r^{-2}$ \cite{Moore:2000fp,Yoshida:2000bx}.} Furthermore, analytical arguments based on self-similar solutions to fluid equations led to similar conclusions \cite{1985ApJS...58...39B}.

\begin{figure*}[t!]
    \centering
    \includegraphics[width=\linewidth]{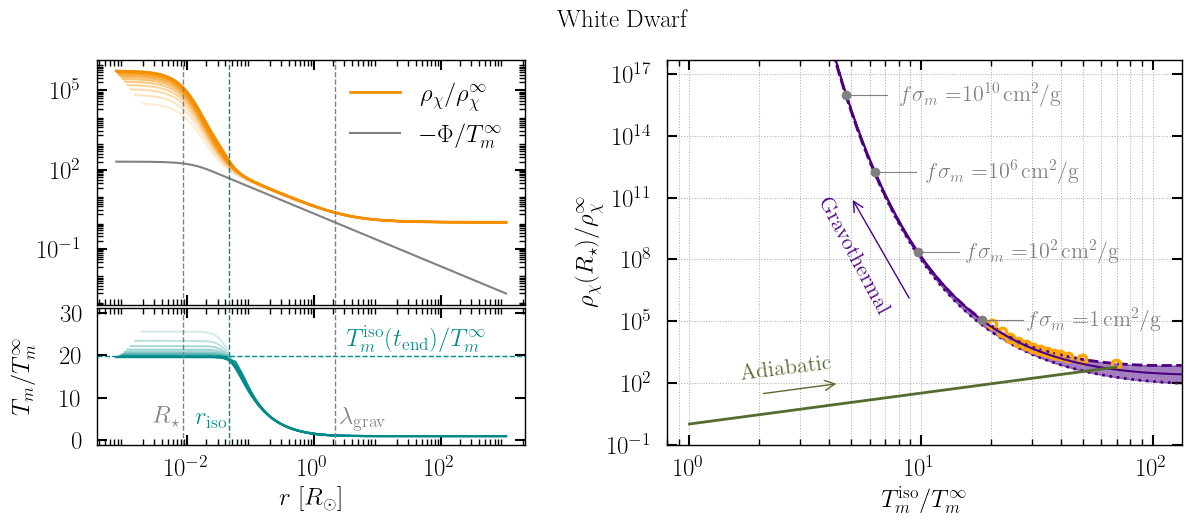}
    \caption{Results of a gravothermal simulation of subcomponent particles $\chi$ accumulating around the benchmark WD of Table~\ref{tab:co_sources}. \textit{Left:} Snapshots of $\chi$'s density $\rho_\chi$, $\chi$'s temperature per unit mass $T_m$, and the WD's gravitational potential $\Phi$, rescaled by the ambient values, $\rho_\chi^\infty$ and $T_m^\infty$, as indicated. The color scale (light to dark) indicates the direction of time evolution. The radius $R_\star$ and gravitational-influence radius $\lambda_{\rm grav}$ of the WD as well as the radius $r_{\rm iso}(t_{\rm end})$ and temperature $T_m^{\rm iso}(t_{\rm end})$ of the isothermal core at the end of the simulation are marked. \textit{Right:} Central density $\rho_\chi(R_\star)$ vs isothermal-core temperature $T_m^{\rm iso}$ phase space, rescaled as indicated. The system first evolves adiabatically to reach the hydrostatic state of Eq.~\eqref{eq:adiabatic}, before proceeding to cool and pile up gravothermally. The orange circles are representative points from the simulation. The purple curves are analytical expectations based on three different assumptions on the $T_m$ profile outside of the isothermal core: (1) the initial adiabatic profile (dashed), (2) the steady-state profile with $\dot{s}\propto -\vec{\nabla}.\vec{\mathcal{F}}\approx 0$ (dotted), and (3) $\eta_{\rm eff}=1$, a simplifying assumption used in the main text (solid). Subcomponents with different $f$ and $\sigma_m$ values evolve along the same evolution curve at rates proportional to $f\sigma_m$ and reach different terminal points after $10\text{ Gyr}$ (the age of the WD), as indicated in gray dots. Apart from that, other features of both plots apply equally for all $f$ and $\sigma_m$ values.}
    \label{fig:profiles}
\end{figure*}

We are interested in the accumulation of a galactic population of $\chi$ particles around a compact object of mass $M_\star$ and radius $R_\star$. To model this process, we employ the \textit{gravothermal} formalism, which was developed to model stellar dynamics in a globular cluster \cite{1980MNRAS.191..483L} and, more recently, applied to SIDM scenarios \cite{Balberg:2002ue,Shapiro:2018vju}. It is based on three equations: (1) $\vec{\nabla}(\rho_\chi T_m)=-\rho_\chi\vec{\nabla}\Phi$, (2) $\rho_\chi T_m\dot{s}=-\vec{\nabla}.\vec{\mathcal{F}}$, (3) $\vec{\mathcal{F}}=-\kappa_{\rm eff}\vec{\nabla}T_m$, which describe, respectively, quasi-hydrostatic equilibrium, the first law of thermodynamics, and Fourier's law of conduction. Here, $s\equiv \ln (T_m^{3/2}/\rho_\chi)$ is the specific entropy of the gas up to additive constants, $\vec{\mathcal{F}}$ is the conductive heat flux, $\kappa_{\rm eff}\sim \rho_\chi^2\sigma_m\sqrt{T_m} (r^{-2}+\lambda_{\rm mfp}^{-2})^{-1}$ is the effective conductivity, and $\lambda_{\rm mfp}\equiv (\rho_\chi\sigma_m)^{-1}$ is the mean free path. We further assume that the system is spherically symmetric and neglect $\chi$'s contribution to the gravitational potential $\Phi$. For more details, see the Appendix~\ref{app:gravothermalformalism}.

Important timescales of our setup include: the collisional timescale $\tau_{\rm col}\equiv \lambda_{\rm mfp}/\sqrt{T_m}$. the sound-crossing timescale $\tau_{\rm sound}\equiv r/\sqrt{T_m}$, and the heat-conduction timescale $\tau_{\rm cond}\equiv |\dot{s}|^{-1}\sim \tau_{\rm col}(1+r^2/\lambda_{\rm mfp}^2)$. The gravothermal approach implicitly assumes that hydrostatic and local thermal equilibrium are rapidly established. These are justified if both $\tau_{\rm col}$ and $\tau_{\rm sound}$ are shorter than $\tau_{\rm cond}$. We checked that these requirements are always satisfied.\footnote{When $\lambda_{\rm mfp}\gtrsim r$, the local-thermalization condition, $\tau_{\rm col}\lesssim\tau_{\rm cond}$, holds only marginally. Nevertheless, the results of gravothermal (with proper calibration of the conductivity coefficients) and N-body simulations have been shown to agree well even in those marginal cases \cite{Ahn:2004xt,2011MNRAS.415.1125K,Yang:2022zkd,2025arXiv250413004M}. } The longest timescale we will consider is $t_{\rm age}= 10\text{ Gyr}$, thus, as a minimum requirement, we assume that the ambient $\chi$ gas has $\tau_{\rm col}\ll t_{\rm age}$, which amounts to $f\sigma_m\gg 1\text{ cm}^2/\text{g}$. Furthermore, we assume that within a timescale $\tau_{\rm sound}\ll \tau_{\rm cond}$, the subcomponent gas rearranges itself isentropically to establish a hydrostatic equilibrium around the compact object; see Appendix.~\ref{app:preGravothermal}. The resulting configuration can be found from  Eq.~\eqref{eq:hydrostatic}, after equating the specific entropy to that of the mean background $s_\chi^\infty=\ln[(T_m^\infty)^{3/2}/\rho_\chi^\infty]$, to be
\begin{align}
    T_m^{\rm ad}(r)=T_m^{\infty}-\frac{2}{5}\Phi(r),\,\,\,\,  
    \rho_\chi^{\rm ad}(r)=\rho_\chi^{\infty}\left(\frac{T_m^{\rm ad}(r)}{T_m^{\infty}}\right)^{3/2}\label{eq:adiabatic}
\end{align}
which we refer to as the adiabatic temperature and density profiles. These serve as an initial condition for the subsequent gravothermal evolution.\footnote{Although the temperature profile of Eq.~\eqref{eq:adiabatic} is consistent with the boundary condition $T_m(r\rightarrow \infty)=T_m^\infty$, it behaves as $T_m^{\rm ad}-T_m^\infty\propto r^{-1}$ beyond the radius of gravitational influence of the compact object $\lambda_{\rm grav}\equiv GM_\star/T_m^\infty$. Thus, such a profile leads to an unphysical heat-conduction luminosity that is growing as $L\equiv 4\pi r^2\mathcal{F}\propto r^2$  at $r\gtrsim \lambda_{\rm grav}$. The details of the small deviation $T_m-T_m^\infty$ are important because $L$ depends on $\partial_r T_m$ and not just $T_m$. In reality, collisions of the $\chi$ gas bound to the compact object with the ambient, unbound gas (which are important only at $r\gtrsim \lambda_{\rm grav}$) would bring $T_m$ exponentially close to $T_m^\infty$ after several $\tau_{\rm col}$. To account for this effect, in practice we introduce an exponential factor to the $T_m^{\rm ad}$ profile to make it decay faster: $T_m^{\rm ad}=T_m^{\infty}-(2/5)\Phi e^{-r^2/\lambda_{\rm grav}^2}$.} 

\section{Gravothermal Pile-Up}
Here, we discuss the subsequent buildup dynamics of $\chi$ particles around gravitational wells. We base our analysis on the results of gravothermal simulations we ran using the code developed by \cite{Pollack:2014rja} and re-implemented by \cite{Nishikawa:2019lsc, Outmezguine:2022bhq, Gad-Nasr:2023gvf}. We modified the codebase provided in \cite{repo_GravothermalSIDM} to adapt it to our setup, with a major update of switching from a self-gravitating system to that with fixed external gravitational potential. This code discretizes the subcomponent gas into concentric spherical shells, and evolves the shells by alternating between (1) a heat conduction step for each shell that brings the system out of hydrostatic equilibrium and (2) isentropic repositioning of shells to reestablish hydrostatic equilibrium. More details about the code are provided in the Appendix~\ref{app:code}. 

In the $\lambda_{\rm mfp}\gg r$ limit, the gravothermal evolution equations are invariant under simultaneous rescalings of $f\sigma_m\rightarrow \alpha\times f\sigma_m$ and $t\rightarrow \alpha^{-1}\times t$. Thus, subcomponents with different $f\sigma_m$ values go through the same series of movie frames (spatial profiles) with rates proportional to $f\sigma_m$. Formally, the solutions for the density and temperature profiles can be written as $\rho_\chi=\rho_\chi^\infty \mathcal{T}_\rho(f\sigma_mt,r)$ and $T_m=T_m^\infty \mathcal{T}_{T}(f\sigma_mt,r)$, where $\mathcal{T}_{\rho}$ and $\mathcal{T}_{T}$ are transfer functions with the indicated dependencies. Note that $\rho_\chi$ carries an additional $f$ dependence through the boundary condition $\rho_\chi^\infty \propto f$ that is not captured by $\mathcal{T}_\rho$. Regardless, the evolution of $\rho_\chi / \rho_\chi^\infty$ and $T_m / T_m^\infty$ depends on $f$ and $\sigma_m$ only through the combination $f \sigma_m t$.

We performed gravothermal simulations for the accumulation of subcomponents around WD up to $t_{\rm end}=0.8\text{ Gyr}[f\sigma_m/(\text{cm}^2/\text{g})]^{-1}$. We plot in Fig.~\ref{fig:profiles} snapshots of the simulated temperature and density profiles. From these results, we could infer that, almost immediately, heat conduction thermalizes the $\chi$ gas near the center, forming an isothermal core. The adiabatic profile outside the core remains essentially frozen at first. Subsequently, the isothermal core cools down by transferring heat to its vicinity, growing in size and mass as a result. The profile exterior to the core develops inside out: heat conduction thaws the profile as its effect spreads to larger radii, and the thawed parts seem to evolve toward a stationary shape characterized by the lack of heat sink/source, $\dot{s}\propto -\vec{\nabla}.\vec{\mathcal F}=0$, whereupon the profile, again, freezes. We note that a similar tendency is found in other systems \cite{Shapiro:2018vju,Shapiro:2014oha,Amaro-Seoane:2004deo,Bahcall:1976aa}. We did not simulate an NS since it is significantly more expensive computationally, but expect the results to be qualitatively the same as that for a WD.\footnote{As a check, we run simulations with hypothetical compact objects with mass $M_\star=1M_\odot$ and radii $R_\star=600\text{ km}$ ($0.1$ times WD radius) and $R_\star=6\times 10^4\text{ km}$ (10 times WD radius), and found that all the results are in agreement with our understanding of the WD case.}

The simulation eventually becomes prohibitively expensive as the timestep required to keep it reliable becomes extremely small. While we have simulation results up to $t=0.8 \text{ Gyr}$ for $f\sigma_m\sim 1\text{ cm}^2/\text{g}$, in much of the interesting $(f,\sigma_m)$ parameter space, namely those with $f\sigma_m\gg 1\text{ cm}^2/\text{g}$, a 10 Gyr run time is beyond our practical reach. In order to determine the final configurations of the $\chi$-particle piles around compact objects, we develop an analytical model of the gravothermal evolution that captures salient aspects of our simulation results. Our simulations suggest that, in general, the temperature profile can be divided into two regions: the isothermal core and the exterior. We approximate the temperature profile inside the isothermal core as exactly uniform and write the full profile as\begin{align}
    T_m(r)&=T_m^{\rm iso}\Theta(r_{\rm iso}-r)+T_m^{\rm ext}(r)\Theta(r-r_{\rm iso})
\end{align}
where $\Theta$ is the Heaviside step function. It suffices to specify the temperature profile, $T_m(r)$, as other properties of a $\chi$ pile can be derived from it. We denote the corresponding hydrostatic density profiles, given by Eq.~\eqref{eq:hydrostatic}, as $\rho_{\chi}^{\rm iso/ext}$. Given a $T_m^{\rm ext}$, we define $r_{\rm iso}$ precisely as the radius at which $\partial_r T_m^{\rm ext}=(T_m^\infty-T_m^{\rm iso})/r_{\rm iso}$. Using the above temperature profile, we are able to reduce the three gravothermal partial differential equations into a single integro-differential equation for the time evolution of $T_m^{\rm iso}$
\begin{align}
    \dot{T}_m^{\rm iso}&\approx   \frac{[r^2\kappa_{\rm eff}^{\rm ext}\partial_rT_m^{\rm ext}]_{r=r_{\rm iso}}}{\int_0^{r_{\rm iso}}r^2 dr\rho_\chi^{\rm iso}\left(\frac{3}{2}-\frac{\partial\ln\rho_\chi^{\rm iso}}{\partial\ln T_m^{\rm iso}}\right)} \label{eq:simpheatEq}
\end{align}
where $\kappa_{\rm eff}\approx r^2\rho_\chi^2\sigma_m\sqrt{T_m}$ for $\lambda_{\rm mfp}\gg r$, which is satisfied in all the parameter space we consider at $r=r_{\rm iso}$. We assume the initial condition $T_m^{\rm iso}(t=0)=T_m^{\rm ad}(r=R_\star)$, although the final results are not sensitive to this choice. Eq.~\eqref{eq:simpheatEq} can be solved numerically once the exterior temperature profile $T_m^{\rm ext}$ is specified, which we discuss next.

The density profile inside the isothermal core is given by Eq.~\eqref{eq:hydrostatic} and can be expressed as
\begin{align}
    \frac{\rho_\chi^{\rm iso}(r)}{\rho_\chi^{\infty}}&=(2\eta_{\rm iso})^{2\bar{\eta}}\left(\frac{T_m^{\rm iso}}{T_m^\infty}\right)^{2\bar{\eta}-1}e^{-\frac{\Phi(r)}{T_m^{\rm iso}}-2\eta_{\rm iso}},\,\, \eta\equiv\frac{r\partial_r\Phi}{2T_m^{\rm ext}} \label{eq:rhoiso}
\end{align}
where $\bar{\eta}\equiv[\int_{r_{\rm iso}}^{\infty} \frac{dr}{r}\, \eta]/[\ln(\lambda_{\rm grav}/r_{\rm iso})]$, $\eta_{\rm iso}\equiv \eta(r_{\rm iso})$, and $\lambda_{\rm grav}\equiv GM_\star/T_m^\infty$ is the compact object's radius of gravitational influence. The quantity $\eta$ encapsulates aspects of the external profiles that are relevant to our analysis. It also measures the conformity of $T_m^{\rm ext}$ with the virial theorem, which predicts $\left<r\partial_r\Phi\right>/2T_m^{\rm ext}=1$, for a particle in potential $\Phi$ in 1D. Indeed, $\eta$ in general does not deviate much from unity. In our simulations, the $\eta$ profile starts with that corresponding to the adiabatic profile of Eq.~\ref{eq:adiabatic} and evolves toward the $\eta$ of the steady temperature profile $T_m^{\rm st}\approx T_m^\infty+(4GM_\star/7r)e^{-7r/4\lambda_{\rm grav}}$ obtained from the stationary condition $\dot{s}\propto -\vec{\nabla}.\vec{\mathcal{F}}\approx0$ and Eq.~\ref{eq:hydrostatic}. Throughout the evolution, the $\eta_{\rm iso}$ and $\bar{\eta}$ remain close to unity. See the Appendix~\ref{app:late} for more details. In Fig.~\ref{fig:profiles}, we display the  $\rho_\chi^{\rm iso}$ at $r=R_\star$ as a function of $T_m^{\rm iso}$, for different $\eta$ assumptions. It is apparent that the different $\eta$ assumptions amount to mild $\mathcal{O}(1)$ variation in $\rho_\chi(R_\star)$ and do not significantly affect the overall trend, especially at late times.

\begin{figure*}
    \centering
    \includegraphics[width=\linewidth]{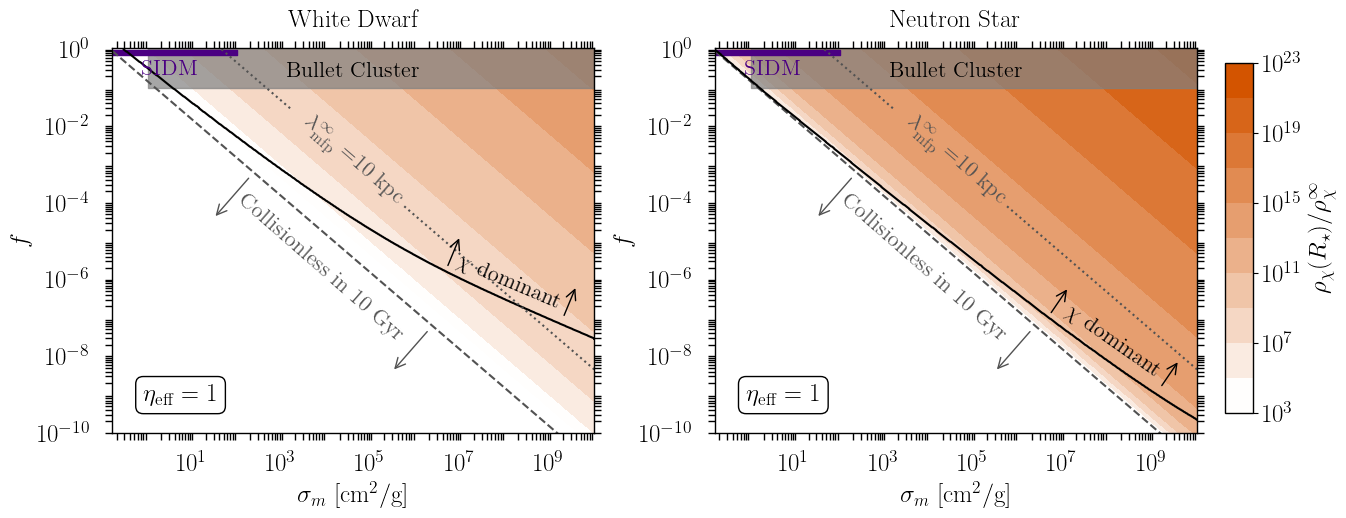}
    \caption{ Subcomponent $\chi$ parameter space. Here, $f$ is $\chi$'s galactic DM fraction, $\sigma_m$ is $\chi$'s cross-section-to-mass ratio, and $\rho_\chi(R_\star)/\rho_\chi^\infty$ is $\chi$'s  final density-enhancement factor at the radius $R_\star$ of the central compact object: white dwarf (\textit{left}) and neutron star (\textit{right}). Regions of the parameter space ruled out by Bullet Cluster observations and favored by the SIDM paradigm are shown in gray and purple, respectively. Below the dashed lines, $\tau_{\rm col}\gtrsim 10\text{ Gyr}$, and so the assumptions of the gravothermal formalism are not satisfied. Above the dotted lines, the ambient subcomponent behaves as a perfect fluid on galactic scales. The $\rho_\chi(R_\star)/\rho_\chi^\infty$ are obtained by integrating Eq.~\eqref{eq:simpheatEq} using Eq.~\eqref{eq:rhoiso} with $\eta_{\rm iso}=\bar{\eta}=\eta_{\rm eff}=1$. Above the solid lines, the final central densities of the accumulated $\chi$ particles satisfy  $\rho_\chi(R_\star)/(0.4\GeV/\text{cm}^3)\gtrsim 20\text{ (WD)}, 600\text{ (NS)}$, and thereby locally dominate over gravitationally focused cold collisionless DM inside the compact objects.}
    \label{fig:parameterspace}
\end{figure*}

To simplify our analysis, in what follows we will set $\eta_{\rm iso}$ and $\bar{\eta}$ to an effective constant $\eta_{\rm eff}$; a more complete account of $\eta$ is given in the Appendix~\ref{app:late}. We solved the heat equation numerically and found that the isothermal+exterior model described above reproduces in detail our simulation results, where they overlap. As cooling of $T_m^{\rm iso}$ continues, increased $\rho_\chi^{\rm iso}$ slows down further cooling, unlike the catastrophic collapse found in SIDM scenarios. The resulting density enhancement depends on how much the $T_m^{\rm iso}$ can cool over the age of the compact object $t_{\rm age}$. We plot in Fig.~\ref{fig:parameterspace} the final enhancement factor $\rho_\chi(R_\star)/\rho_\chi^\infty$ in the compact objects of Table~\ref{tab:co_sources} as a function of $f$ and $\sigma_m$, for $\eta_{\rm eff}=1$. In the same figure, we delineate the regime where the subcomponent $\chi$ becomes the dominant DM, locally, inside the compact object, accounting for the fact that even the dominant, collisionless DM has an enhanced density due to gravitational focusing. By Liouville theorem, we can infer that $\rho_{\rm CDM}(R_\star)/\rho_{\rm CDM}^\infty\sim v_{\rm esc}/\sqrt{T_m^\infty}$, which amounts to about 20(600) for WDs(NSs). 

Next, we provide an analytical estimate for the order of magnitude of the final $\rho_\chi(R_\star)$. Very crudely, we can approximate the numerator of the right hand side of Eq.~\eqref{eq:simpheatEq} as $-\sigma_mr_{\rm iso}^3(\rho_\chi^\infty)^2\left(\frac{T_m^{\rm iso}}{T_m^\infty}\right)^{4\eta_{\rm eff}-2}  (T_m^{\rm iso})^{3/2}$ and the denominator as  $R_\star^3\rho_\chi^\infty \left(\frac{T_m^{\rm iso}}{T_m^\infty}\right)^{2\eta_{\rm eff}-1} e^{\frac{v_{\rm esc}^2/2}{T_m^{\rm iso}}}\left(\frac{v_{\rm esc}^2/2}{T_m^{\rm iso}}\right)$, assuming $\frac{\partial\ln\rho_\chi^{\rm iso}}{\partial\ln T_m^{\rm iso}}\approx -\frac{\Phi}{T_m^{\rm iso}}\sim \frac{v_{\rm esc}^2/2}{T_m^{\rm iso}}\gg 3/2$ and the contributions from $r\sim R_\star$ dominate the integral. The heat equation then reduces to
\begin{align}
    \frac{\dot{T}_m^{\rm iso}}{T_m^{\rm iso}}&\sim \frac{-1}{\tau_{\rm col}^\infty}\left(\frac{v_{\rm esc}^2}{2T_m^\infty}\right)^{2\eta_{\rm eff}-\frac{1}{2}}\left(\frac{v_{\rm esc}^2}{2T_m^{\rm iso}}\right)^{\frac{5}{2}-2\eta_{\rm eff}}e^{-\frac{v_{\rm esc}^2}{2T_m^{\rm iso}}}
\end{align}
where $\tau_{\rm col}^\infty=(\rho_\chi^\infty\sigma_m\sqrt{T_m^\infty})^{-1}$. The final $T_m^{\rm iso}$ can be estimated by equating the above with $-t_{\rm age}^{-1}$, and solving  the resulting equation iteratively, giving $(v_{\rm esc}^2/2)/T_{m}^{\rm iso}|_{t=t_{\rm age}}\approx \ln x+(5/2-2\eta_{\rm eff})\ln(\ln x)$, with $x=t_{\rm age}/\tau_{\rm col}^\infty$. For $\eta_{\rm eff}=1$, this yields
\begin{align}
    \frac{v_{\rm esc}^2/2}{T_m^{\rm iso}|_{t=t_{\rm age}}}
    &\approx 20.4+3\ln\left(\frac{v_{\rm esc}}{0.02}\right)+\ln\left(\frac{f\sigma_m}{10^4\text{cm}^2/
    \text{g}}\right)\nonumber\\
    &=\begin{cases}
       20.4+\ln\left(\frac{f\sigma_m}{10^4\text{ cm}^2/\text{g}}\right) &\text{(WD)}\\
       30.6+\ln\left(\frac{f\sigma_m}{10^4\text{ cm}^2/\text{g}}\right) &\text{(NS)}
    \end{cases}
\end{align}
The corresponding ratio of the central density to the ambient density, given by Eq.~\ref{eq:rhoiso}, is
\begin{align}
    \frac{\rho_\chi(R_\star)|_{t=t_{\rm age}}}{\rho_{\chi}^{\infty}}
    &\sim 1\times 10^9\left(\frac{v_{\rm esc}}{0.02}\right)^{5}\left(\frac{f\sigma_m}{10^{4}\text{ cm}^2/\text{g}}\right)\nonumber\\
    &=\begin{cases}
        1\times 10^9\left(\frac{f\sigma_m}{10^{4}\text{ cm}^2/\text{g}}\right) &\text{(WD)}\\
        2\times 10^{16}\left(\frac{f\sigma_m}{10^{4}\text{ cm}^2/\text{g}}\right) &\text{(NS)}
    \end{cases}
\end{align}
These rough estimates are in agreement with the previously obtained results, displayed in Fig.~\ref{fig:parameterspace}, based on numerically solving Eq.~\eqref{eq:simpheatEq}. Since $\rho_\chi^\infty\propto f$, this translates to the final central density scaling as $\rho_\chi(R_\star)|_{t=t_{\rm age}}\propto f^2\sigma_m$. Hence, in scenarios with multiple DM components, even those with extremely tiny $f$'s can accumulate to considerable levels if they have compensatingly large $\sigma_m\propto f^{-2}$.

The dependence of the final central density on the compact object enters through its $\rho_\chi(R_\star)|_{t=t_{\rm age}}\propto v_{\rm esc}^5 t_{\rm age}$ scaling, which explains how the $\rho_\chi(R_\star)$ for an NS is a factor of $\sim (0.6/0.02)^5=2\times 10^7$ higher than that for a WD. The total captured mass is concentrated at $r\lesssim R_\star$ where $\rho_\chi$ is exponentially enhanced. It can be estimated as $M_{\rm iso}\sim (4\pi/3) R_\star^3\rho_\chi(R_\star)\propto v_{\rm esc}^5R_\star^3t_{\rm age}$. Compared to a WD, an NS has a much higher $v_{\rm esc}$ but much lower $R_\star^{3}$. For the benchmark properties listed in Table~\ref{tab:co_sources}, this amounts to $M_{\rm iso}$ that is one order of magnitude smaller in an NS than in a WD. To further quantify the efficiency of gravothermal pile-up, we can also compute its effective capture radius $R_{\rm cap}$, defined by $\pi R_{\rm cap}^2=M_{\rm iso}/(\rho_\chi^\infty \sqrt{T_m^\infty}t_{\rm age})$. As an example, fixing $f=10\%$ and $\sigma_m=10^{10}\text{ cm}^2/\text{g}$, we find that for WDs(NSs) $M_{\rm iso}\sim 10^{-17} M_\odot$($10^{-18} M_\odot$) and $R_{\rm cap}\sim 6000\text{ km}$ (500 km). The obtained $M_{\rm iso}$'s are much larger than the total mass of the initial adiabatic cloud, which we estimate to be $M_{\rm ad}\approx 4\times 10^{-22}fM_\odot$. The $R_{\rm cap}$ values for the chosen parameters are about the geometrical radius $R_\star$ for WDs and far exceeds $R_\star$ for NSs. 

\section{Discussion}
We have shown how dark particles with a large, elastic, self-scattering cross section comprising a tiny fraction of the dark matter in an environment with a velocity dispersion $\sigma_v$ accumulate in gravitational wells with escape velocities $v_{\rm esc}\gg \sigma_v$. The minimality of the requisite conditions for this process to occur suggests its ubiquitous relevance.

The $\sigma_m$ values assumed in this study arise in many familiar particle-physics models. For instance, a scalar singlet \cite{McDonald:2001vt,Heikinheimo:2016yds,Hochberg:2014dra,Burgess:2000yq,Bernal:2017kxu,Chang:2022psj} has a self-scattering $\sigma_m=9\lambda^2/32\pi m^3\sim 10^4\lambda^2\text{ cm}^2/\text{g}(m/\MeV)^{-3}$ that could be large for a perturbative quartic coupling $\lambda$ if the mass $m$ is small. In models of composite dark particles, including the dark analogs of glueballs \cite{Boddy:2014yra,Jo:2020ggs,Soni:2016gzf}, mesonic \cite{Bhattacharya:2013kma}, baryonic, atomic \cite{Kaplan:2009de,Cline:2013pca,Cyr-Racine:2012tfp,Boddy:2016bbu}, and molecular \cite{Cline:2013pca,Ryan:2021dis} states \cite{Cline:2013zca,Kribs:2016cew}, large elastic $\sigma_m$'s arise naturally as a consequence of the particles' residual strong self-interactions, large geometric sizes, or the presence of light force-mediators. If the binding energy of one of these states is much greater than their typical kinetic energy, their collisions are mostly elastic. We discuss the simplest of these models, scalar singlet and dark atom, further in the Appendix~\ref{app:model}. Some of these models predict velocity-dependent cross-sections \cite{Kaplinghat:2015aga,Agrawal:2016quu,Outmezguine:2022bhq}. Our analysis could be straightforwardly generalized to such cases as well.  

Apart from WDs and NSs in galaxies, many other combinations of small $\sigma_v$ and large $v_{\rm esc}$ exist. Small velocity dispersions $\sigma_v\ll 10^{-3}$ can be found in Galaxy outskirts, dwarf galaxies \footnote{Subcomponent particles with sufficiently small mean free paths can be shielded from ram stripping in dwarf galaxies. The fact that gas-rich dwarf galaxies exist suggests that this outcome is possible \cite{2021ApJ...913...53P,2009ApJ...696..385G}.}, or scenarios where the subcomponent forms a dark disk via inelastic processes \cite{Fan:2013yva,Fan:2013tia}. Other gravitational wells with $v_{\rm esc}\gtrsim 10^{-3}$ include massive main-sequence (MS) stars and population III stars.\footnote{Black holes have $v_{\rm esc}=1$ but their absorptive inner boundary conditions act as a central sink that hinders pile-up.} Based on the fitting formulas in the Appendix of Ref.~\cite{Nguyen:2023czp}, the radius of MS star scales with its mass $M_\star$ as $R_\star\propto M_\star^{0.6}$, which translates to $v_{\rm esc}^2 \propto M_\star^{0.4}$. Long-range dark forces could deepen the potential well felt by $\chi$s relative to gravity \cite{Bogorad:2024hfj,Graham:2025gtd}, potentially leading to substantial $\chi$ piles even around the Sun or Earth. Similar piles may also form around macroscopic DM states \cite{Jacobs:2014yca,Bai:2020jfm,Tolos:2015qra,Grabowska:2018lnd,Kaplan:2024dsn,Fedderke:2024hfy,Ebadi:2021cte}. 

We have found that within the parameter space we considered, $f\lesssim 10\%$ and $\sigma_m\lesssim 10^{10}\text{cm}^2/\text{g}$, a total captured mass of up to $10^{-17}M_\odot$($10^{-18}M_\odot$) for a WD(NS) in a typical galaxy is possible. Bigger $\chi$ piles can be achieved with $\sigma_m\gtrsim 10^{10}\text{ cm}^2/\text{g}$  and with more favorable combinations of $R_\star$, $v_{\rm esc}$, $T_m^\infty$, and $\rho_\chi^\infty$, although we expect the total captured $\chi$ mass to be limited by the incoming $\chi$ flux to be less than $M_{\rm max}=\rho_\chi^{\infty} \sqrt{T_m^\infty} \pi\lambda_{\rm grav}^2\times 10\text{ Gyr}\sim 10^{-7}f M_\odot$ ($10^{-3}fM_\odot$) for a solar-mass object in a galaxy (dwarf galaxy). Measurably altering observable properties of compact objects via gravity typically requires percent-level or higher DM mass fraction \cite{Bertone:2007ae,Deliyergiyev:2019vti,Sandin:2008db,Ellis:2017jgp,Koehn:2024gal,Karkevandi:2021ygv,Leung:2022wcf,Giangrandi:2022wht}. Piles of $\chi$ around massive MS stars in dwarf galaxies, for instance, might be within the realm of detectability even in the simplest case with only gravity and $\chi$ self-scatterings.

In some cases, catastrophic outcomes are possible with much smaller captured mass. Depending on the model, when $\chi$ accumulates to sufficiently high central densities, various inelastic \cite{Pearce:2013ola,Cirelli:2016rnw,Pearce:2015zca,Freese:2015mta}, number-changing \cite{Bell:2013xk,Hochberg:2014dra,Pappadopulo:2016pkp,Ralegankar:2024zjd}), or quantum-statistics \cite{Ralegankar:2024zjd,Budker:2023sex,Eby:2015hsq,Bramante:2013hn,Kouvaris:2011fi,McDermott:2011jp,Bertoni:2013bsa} effects may turn on. The pile may also become self-gravitating \cite{Goldman:1989nd,Gould:1989gw,Bell:2013xk,Bramante:2013hn,Bramante:2017ulk,Dasgupta:2020mqg,McDermott:2011jp,Garani:2018kkd}.
In the presence of $\chi$-baryon coupling, $\chi$ particles in the pile may exchange energy with the compact object through scatterings, potentially causing them to sink even deeper \cite{Guver:2012ba,Bell:2021fye,Dasgupta:2019juq,Dasgupta:2020dik} and causing extra cooling or heating of the compact object \cite{Bertone:2007ae,Kouvaris:2007ay}. Sufficiently light $\chi$ could be sourced by stellar cores, likely with collective dynamics and nontrivial interplay with the captured $\chi$
\cite{Chang:2022gcs,Berlin:2024ewa}. The gravitational accumulation mechanism studied here could be a precursor or a contributing effect to the scenarios studied in \cite{Ralegankar:2024zjd,Budker:2023sex,Wu:2022wzw,Bai:2020jfm, Eby:2015hsq,Bai:2016wpg,Armstrong:2023cis,Bramante:2013hn,Kouvaris:2011fi,McDermott:2011jp}. 

If the $\chi$ particles annihilate into final states that include SM particles escaping the compact object, the resulting indirect-detection signals could be significantly enhanced due to the $n_\chi^2$ dependence of the annihilation rate and the concentrated density of $\chi$ around compact stars \cite{Leane:2021ihh,Nguyen:2022zwb,Linden:2024uph}. Details of the resulting photon flux are, of course, model-dependent. For now, we can compute part of the $J$ factor, $J(r)\propto \int_0^{r}4\pi r^{\prime 2} dr' (\rho_\chi^{\rm iso})^2$, which is independent of the microphysics of $\chi$. In a region where the typical separation between compact objects is $0.1\text{ pc}$, the annihilation signal from $\chi$ inside compact objects is stronger than that of the ambient $\chi$ if the $J$-factor ratio $J(R_\star)/J(1\text{ pc})\sim (R_\star/\text{pc})^3[\rho_\chi(R_\star)/\rho_\chi^\infty]^2$ is greater than unity, which amounts to  $\rho_\chi(R_\star)/\rho_\chi^\infty\gtrsim 10^{13}$ ($10^{17}$) for WDs(NSs). Density enhancements at those levels are achievable for sufficiently large $f\sigma_m$. Decay signals would not have enhanced strengths on average, but would have different morphologies \cite{Agrawal:2017pnb}, if the telescope could resolve them.

Therefore, through gravothermal pile-up even a tiny subcomponent of dark matter can leave observable signatures. We defer detailed investigation of these signals to separate work.

\section*{Acknowledgments}
We thank Abhishek Banerjee for early collaboration, and Peter Graham and David E. Kaplan for useful conversations. R.E. is supported by Grant No. 63034 from the John Templeton Foundation and the University of Maryland Quantum Technology Center. E.H.T. acknowledges support by NSF Grant No. PHY-2310429, Simons Investigator Award No. 824870, the Gordon and Betty Moore Foundation Grant No. GBMF7946, and the U.S. Department of Energy (DOE), Office of Science, National Quantum Information Science Research Centers, Superconducting Quantum Materials and Systems Center (SQMS) under Contract No. DEAC02-07CH11359.

\appendix
\widetext
\section{Gravothermal Fluid Formalism}
\label{app:gravothermalformalism}
Our setup consists of a gravitational potential well and a thermal fluid, denoted here by subscripts `pot' and $\chi$, respectively, to prevent ambiguity in notation when necessary. We focus on a parameter space where the self-gravity of the fluid can be neglected. The gravothermal set of equations can be written as
\begin{align}
    \text{Hydrostatic equilibrium:} && &\partial_r(\rho_\chi T_m)=-\frac{GM_{\rm pot}(r)}{r^2}\rho_\chi \label{eq:gravothermal1}\\
    \text{Energy flux:} && &\frac{L}{4\pi r^2}=-\kappa_{\rm eff} \partial_r T_m \label{eq:gravothermal2}\\
    \text{1st law of thermodynamics:} && &\partial_r L=-4\pi r^2 \rho_\chi T_m\partial_t\left(\ln\frac{T_m^{3/2}}{\rho_\chi}\right) \label{eq:gravothermal3}
\end{align}
where $M_{\rm pot}(r)$ is the mass enclosed within a sphere of radius $r$, $T_m$ is the temperature \textit{per unit mass} of the fluid, and $L$ is the  outward heat-conduction luminosity. 

The effective conductivity $\kappa_{\rm eff}$ captures both the short mean free path (SMFP) and long mean free path (LMFP) regimes, as well as interpolating between the two:
\begin{equation}
    \kappa_{\rm eff}=\left(\kappa_{\rm smfp}^{-1}+\kappa_{\rm lmfp}^{-1}\right)^{-1}
\end{equation}
where
\begin{equation}
    \kappa_{\rm smfp}=\frac{3}{2}\rho_\chi C_{\rm smfp} \frac{\lambda_{\rm mfp}^2}{\tau_{\rm col}}\,,\quad \kappa_{\rm lmfp}=\frac{3}{2}\rho_\chi C_{\rm lmfp} \frac{H^2}{\tau_{\rm col}}
\end{equation}
The numerical factors $C_{\rm smfp}=b/a=25\pi^2/168$ and $C_{\rm lmfp}=0.753$ are $\mathcal{O}(1)$. $C_{\rm smfp}$ can be derived from first principles, whereas $C_{\rm lmfp}$ is obtained by fitting numerical results such that $\kappa_{\rm eff}$ gives acceptable results in the interpolation regime \cite{Ahn:2004xt,2011MNRAS.415.1125K,Yang:2022zkd,2025arXiv250413004M}. In the SMFP regime, the relevant length scale is the distance between successive collisions, $\lambda_{\rm mfp}=1/(\rho_\chi\sigma_m)$. In the LMFP regime, each particle completes multiple orbits between collisions, during which the order of magnitude of the change in its radial position is given by $H\sim v_rt_{\rm orbit}$ \cite{Sabarish:2025hwb}. Here, $v_r\simeq\sqrt{T_m}$ is radial velocity dispersion, and $t_{\rm orbit}$ is orbital timescale. Outside the gravitational potential of a star $v_r\sim\sqrt{GM_\star/r}$ and $t_{\rm orbit}\sim\sqrt{r^3/GM_\star}$, while inside the star $v_r\sim\sqrt{GM_\star (3R_\star^2 -r^2)/2R_\star^3}$ and $t_{\rm orbit}\sim\sqrt{R_\star^3/GM_\star}$. In both cases, we assumed that the virial theorem holds, as justified by the long mean free path ($\lambda_{\rm mfp} \gg H$). We find $H\sim r$ (as in \cite{Shapiro:2014oha,Shapiro:2018vju}) as long as $r$ is not much smaller than $R_\star$. This length scale is consistent with the definition of the gravitational scale height: the characteristic distance over which the thermodynamic properties of the system change significantly due to the gravitational potential, i.e., $H\partial_r\Phi  \sim T_m$. Which also implies, $H\sim \Phi/\partial_r\Phi\sim r$.

\textit{Gravothermal evolution.} In the parameter space relevant to the current work, the system begins in the LMFP regime (characterized by high temperature and low density) and undergoes gravothermal evolution (leading to lower temperatures and higher densities). This evolution naturally drives the system from the LMFP regime to the SMFP regime. In either of the limiting cases, deep LMFP or deep SMFP, one of the corresponding conductivities becomes small and thus dominates the total effective conductivity, resulting in inefficient heat transport. However, near the transition boundary between the two regimes, both conductivities are comparable and larger than in either limit. Consequently, the gravothermal evolution is the fastest near this boundary, a regime we dub the intermediate mean free path (IMFP) regime. The transition happens when
\begin{equation}
    \rho_\chi^{\rm imfp}  \simeq \sqrt{\frac{C_{\rm smfp}}{C_{\rm lmfp}}} \frac{1}{r \sigma_m}
\end{equation}

The gravothermal fluid formalism is based on Fourier's law of thermal conduction, namely Eq.~\eqref{eq:gravothermal2}. Fundamentally, the latter is a concept that is derived from an expansion of the energy-momentum tensor of a near-perfect fluid to first order in the MFP. In order for the expansion to make sense, strictly speaking the MFP must be tiny compared to other length scales of the problem, i.e. in the SMFP region. However, the situation in the LMFP regime is not as severe as it naively seems. In the LMFP case, the effective MFP is set by the gravitational scale height instead of the collisional MFP. The former is typically of the same order as $r$, which sets the fractional gradients of many quantities. So the small MFP expansion is at the very least marginally justified. In fact, it has been shown that if the prefactor of the conductivity is properly calibrated, the fluid formalism can produce results that agree well with that of N-body simulations even in the LMFP limit \cite{Ahn:2004xt,2011MNRAS.415.1125K,Yang:2022zkd,2025arXiv250413004M}.

\section{Hydrodynamical Simulations of Gravothermal Evolution}
\label{app:code}
We use the method developed in \cite{Pollack:2014rja} and adapted by \cite{Nishikawa:2019lsc, Outmezguine:2022bhq, Gad-Nasr:2023gvf} (codebase available at \cite{repo_GravothermalSIDM}) to solve the set of gravothermal fluid equations, Eqs.~\eqref{eq:gravothermal1}, \eqref{eq:gravothermal2},\& \eqref{eq:gravothermal3}. For a description of the numerical implementation, see Appendix A of \cite{Nishikawa:2019lsc}. A high-level summary of the workflow is as follows:
\begin{itemize}
    \item {\it Divide the fluid into spherically symmetric shells:} The fluid is discretized into multiple concentric shells. Each shell has a fixed mass throughout evolution.
    \item {\it Heat conduction step:} Small amount of heat transfer occurs between neighboring shells. The time step used is much shorter than the collisional timescale $\tau_{\rm col}$. This process slightly perturbs hydrostatic equilibrium.
    \item {\it Hydrostatic adjustment steps:} Following the heat conduction, the system is driven back toward hydrostatic equilibrium. This is done through multiple adjustment steps during which the positions, pressures, and densities of the shells are updated iteratively. This process conserves entropy.
\end{itemize}

\subsection{Dimensionless Quantities and Equations}
The gravothermal fluid equations involve a set of six physical quantities: \{$r$, $M_{\rm pot}$, $\rho_\chi$, $T_m$ (or $v_\chi)$, $t$, $\sigma_m$, $L$\}. To convert the gravothermal fluid equations into a dimensionless form (denoted by tilde), four constraints must be imposed on the scaling quantities (denoted by the subscript 0). Note that the energy flux equation contributes two of these constraints due to the presence of both LMFP and SMFP conductivity regimes. This leaves three degrees of freedom to choose independent scaling quantities. We select $\{M_0, r_0, \rho_0\}$. Other parameters of interest can be made dimensionless through suitable combinations of these three scaling quantities:
\begin{align}
    T_0 &= v_0^2 = \frac{GM_0}{r_0} \\
    \sigma_0 &= \frac{1}{\rho_0 r_0} \\
    L_0 &= \frac{4\pi\rho_0v_0^2r_0^3}{t_0} \label{eq:scaleluminosity}\\
    t_0 &= \frac{1}{a\sigma_m\rho_0v_0} \label{eq:t0params} \\
    \Phi_0 &= v_0^2  
\end{align}
where $a=4/\sqrt{\pi}$. Note that $t_0$ is chosen such that $\tilde{\tau}_{\rm col} = (\tilde{\rho}\tilde{v})^{-1}$. Although the gravitational potential $\Phi$ does not appear explicitly in the fluid equations, it is used in setting the adiabatic initial conditions. Therefore, we also define a corresponding, and consistent scaling quantity $\Phi_0$. The dimensionless gravothermal equations are
\begin{align}
    \partial_{\tilde{r}}\left(\tilde{\rho}_\chi \tilde{v}_\chi^2\right) &= -\frac{\tilde{M}_{\rm pot}\tilde{\rho}_\chi}{\tilde{r}^2} \\
    \tilde{L} &= -\frac{3}{2}\tilde{r}^2\tilde{v}_\chi\bigg(\frac{1}{C_{\rm lmfp}}\frac{1}{\tilde{r}^2 \tilde{\rho}_\chi^2} + \frac{\tilde{\sigma}_m^2}{C_{\rm smfp}}\bigg)^{-1} \partial_{\tilde{r}}\tilde{v}_\chi^2 \label{eq:Ltilde_eq}\\
    \partial_{\tilde{r}} \tilde{L} &= -\tilde{r}^2 \tilde{\rho}_\chi \tilde{v}_\chi^2 \partial_{\tilde{t}}\bigg(\ln\frac{\tilde{v}_\chi^3}{\tilde{\rho}_\chi}\bigg) 
\end{align}
Therefore the transition between LMFP and SMFP happens when
\begin{equation}
    \tilde{\rho}_\chi\simeq \sqrt{\frac{C_{\rm smfp}}{C_{\rm lmfp}}}\frac{1}{\tilde{r} \tilde{\sigma}_m}
\end{equation}

\subsection{Gaussian Potential Well and Adiabatic Initial Condition}
We model the gravitational potential well with a Gaussian density profile: $\rho_{\rm pot}=\rho_\star\exp\left(-r^2/2R_\star^2\right)$, which implies a total mass of $M_\star = (2\pi R_\star^2)^{3/2}\rho_\star$. The dimensionless quantities used in our numerical computation are
\begin{align}
    \tilde{M}_{\rm pot}(\tilde{r}) &= \tilde{M}_\star\left[\text{erf}\left(\frac{\tilde{r}}{\sqrt{2}\tilde{R}_\star}\right)-\sqrt{\frac{2}{\pi}}\frac{\tilde{r}}{\tilde{R}_\star}\exp\left(-\tilde{r}^2/2\tilde{R}_\star^2\right)\right]\\
    \frac{{\rm d}\tilde{M}_{\rm pot}(\tilde{r})}{{\rm d}\tilde{r}} &= \tilde{M}_\star\sqrt{\frac{2}{\pi}}\frac{\tilde{r}^2}{\tilde{R}_\star^3}\exp\left(-\tilde{r}^2/2\tilde{R}_\star^2\right)\\
    \tilde{\Phi} &= \tilde{M}_\star\left[-\frac{1}{\tilde{r}}\text{erf}\left(\frac{\tilde{r}}{\sqrt{2}\tilde{R}_\star}\right)\right]
\end{align}

For the adiabatic initial condition, we use 
\begin{align}
    \tilde{T}_m(\tilde{r}) &= \tilde{T}_m^\infty - \frac{2}{5}\tilde{\Phi}(\tilde{r})  e^{-\tilde{r}^2/\tilde{\lambda}_{\rm grav}^2}\\
    \tilde{\rho}(\tilde{r}) &= \tilde{\rho}_\chi^\infty\frac{\tilde{T}_m^\infty}{\tilde{T}_m(\tilde{r})}e^{-\int_{\infty}^r dr'\, \frac{\partial_{r'}\Phi(r')}{T_m(r')}}
\end{align}

\subsection{Choices for Scaling Quantities and Mapping Simulation Input Parameters to Physical Parameters}
We make the following choices, which determine the physical interpretation of the input dimensionless parameters:
\begin{align}
    M_0 &= M_\star \\
    \rho_0 &= \rho_{\rm CDM} \label{eq:scalerho}\\
    r_0 &= \frac{GM_0}{v_{\infty}^2}
\end{align}
Given these three choices, we compute the rest of the scaling quantities
\begin{align}
    v_0 &= \sqrt{GM_0/r_0} = v_{\infty} \label{eq:scalev}\\
    t_0 &= 6.6\times10^{-11}\,{\rm Gyr}\left(\frac{10^{10}\,{\rm cm^2/g}}{\sigma_m}\right)\left(\frac{0.4\,{\rm GeV/cm^3}}{\rho_{\rm CDM}}\right)\left(\frac{10^{-3}}{v_\infty}\right)\label{eq:scalet} \\
    \sigma_0 &= 9.5\times10^{12} \, {\rm cm^2/g} \left(\frac{0.4\,{\rm GeV/cm^3}}{\rho_{\rm CDM}}\right)\left(\frac{M_\odot}{M_\star}\right)\left(\frac{v_\infty}{10^{-3}}\right)^2\label{eq:scalesigma}
\end{align}
Given these scaling quantities, we can compute dimensionless parameters 
\begin{align}
    \tilde{M}_\star &= M_\star/M_0 = 1 \\
    \tilde{\lambda}_{\rm grav} & = (GM_\star/T_m^\infty)/r_0 = 1 \\
    \tilde{\rho}_\chi^\infty &= \rho_\chi^{\infty}/\rho_{\rm CDM} = f_\chi \\
    \tilde{R}_\star & = R_\star/r_0 = 2v_0^2/v_{\rm esc}^2 = 2\tilde{v}_{\rm esc}^{-2}
\end{align}

\begin{table}[t!]
\small
\caption{Benchmark properties of compact objects.} \label{tab:co_sources_sm}
\begin{center}
\begin{tabular}{l|c|c|c|c|c|c|c}\hline\hline
\begin{tabular}{@{}c@{}}{}\\ \end{tabular} & 
\begin{tabular}{@{}c@{}}$M_{\star}\,[M_\odot]$ \\ \end{tabular} & 
\begin{tabular}{@{}c@{}}$R_{\star}\,[\text{km}]$ \\ \end{tabular} &
\begin{tabular}{@{}c@{}}$r_0\,[R_\odot]$ \\ \end{tabular} &
\begin{tabular}{@{}c@{}}$v_{\rm esc}$ \\ \end{tabular} &
\begin{tabular}{@{}c@{}}$\tilde{M}_{\star}$ \\ \end{tabular} & 
\begin{tabular}{@{}c@{}}$\tilde{R}_{\star}$ \\ \end{tabular} &
\begin{tabular}{@{}c@{}}$\tilde{v}_{\rm esc}$ \\ \end{tabular}\\ \hline\hline
{} & {} & {} & {} & {} & {} & {} \\
white dwarf &$1$ & $6000$ & $2.12$ & $0.02$ & 1 & $4.0\times10^{-3}$ & 20\\
neutron star &$1.4$ & $10$ &$2.97$ & $0.6$ & 1 &  $4.8\times10^{-6}$ & 600\\
{} & {} & {} & {} & {} & {} & {}\\

\hline
\end{tabular}
\end{center}
\end{table}

\subsection{Modifications to the Original Codebase of \cite{repo_GravothermalSIDM}}
We adapt the original code base with the following modifications to make it suitable for the scenarios studied in this work:
\begin{itemize}
    \item The original code base models self-gravitating systems; we modify it to use a fixed gravitational potential that is independent of the dark matter distribution. In other words, we neglect the gravitational influence of the dark matter relative to that of the central compact object.
    \item We update the gravitational scale height, and consequently the effective conductivity. In the original code base, the scale height is defined as $H=\sqrt{T_m/4\pi G\rho_\chi}$, which follows from the assumption of a self-gravitating system. In our case, as justified above, we instead set $H=r$.
    \item As explained in Appendix A of \cite{Nishikawa:2019lsc}, hydrostatic equilibrium is achieved through a linearized hydrostatic condition, expressed in the form of a tridiagonal equation $a_i\, \Delta\tilde{r}_{i-1} + b_i\, \Delta\tilde{r}_{i} + c_i\, \Delta\tilde{r}_{i+1} = d_i.$ We need to modify the coefficients \(a_i\), \(b_i\), \(c_i\), and \(d_i\) to account for the fact that the external gravitational potential (and its corresponding mass distribution) is fixed.  As a consequence, the mass enclosed within the \(i\)th shell changes as the shells move, unlike the self-gravitating case where each shell is defined with a fixed mass and the enclosed mass remains constant throughout the simulation. To account for this effect, we apply the following modification proportional to $d \tilde{M}_\star(\tilde{r})/d\tilde{r}$ in the linearized hydrostatic condition equation:
    \begin{equation}
        b_i = b_i \bigg|_{\rm original~codebase}-\frac{1}{4} \dfrac{d \tilde{M}_\star(\tilde{r})}{d\tilde{r}}\bigg|_{\tilde{r}=\tilde{r}_i} (\tilde{\rho}_i + \tilde{\rho}_{i+1})(\tilde{r}_{i-1} - \tilde{r}_{i+1})
    \end{equation}
    which we derive in the next subsection.     Without this additional linear term, the hydrostatic adjustment step following each heat conduction iteration converges slowly--if at all--in the simulations.
    \item Eq.\,\ref{eq:Ltilde_eq} is consistent with Eq.\,(A1d) of \cite{Nishikawa:2019lsc} (taking into account the modified scale height in our work). However, the original codebase \cite{repo_GravothermalSIDM} adopts a different choice of scaling time, $ t_0^{\rm code} = 1/\sqrt{4\pi \rho_0 G}$. This implies $ t_0^{\rm code}/t_0^{\rm paper} = a \, \tilde{\sigma}_m$. As a result, the definition of $\kappa_{\rm eff}$ in the code differs from that in the paper, namely $\kappa_{\rm code}/a \sigma_m = \kappa_{\rm paper}$.  We emphasize that despite these different choices for $t_0$, the implementation remains consistent.  We implement Eq.\,\ref{eq:Ltilde_eq} in the code using the definition of $t_0$ given in Eq.\,\ref{eq:t0params}.
\end{itemize}

\subsection{Linearized Hydrostatic Equilibrium Equation} 
In this subsection, we omit the tildes on all physical parameters for simplicity, noting that all parameters are indeed the dimensionless, tilde versions. Let us compute the $\{a_i,b_i,c_i,d_i\}$ coefficients for the hydrostatic equilibrium equation mentioned above. The perturbed hydrostatic equilibrium equation is
\begin{align}
    \frac{p_{i+1}+\Delta p_{i+1}-p_i-\Delta p_i}{\left(r_{i+1}+\Delta r_{i+1}-r_{i-1}-\Delta r_{i-1}\right)/2}+\frac{\rho_i+\Delta \rho_i+\rho_{i+1}+\Delta \rho_{i+1}}{2}\frac{M_{\star,i}+\Delta M_{\star,i}}{r_i^2+2r_i\Delta r_i}=0
\end{align}
where
\begin{align}
    \Delta \rho_i&=-3\rho_i\frac{r_i^2\Delta r_i-r_{i-1}^2\Delta r_{i-1}}{r_i^3-r_{i-1}^3}\\
    \Delta p_i&=-5p_i\frac{r_i^2\Delta r_i-r_{i-1}^2\Delta r_{i-1}}{r_i^3-r_{i-1}^3}\\
    \Delta M_{\star,i}&=\frac{dM_{\star}}{dr}\bigg|_{r_i}\Delta r_i
\end{align}
Note that $\Delta M_{\star,i}$ is the new term which is absent in the self-gravitating systems considered before. At linear order we have $a_i\, \Delta\tilde{r}_{i-1} + b_i\, \Delta\tilde{r}_{i} + c_i\, \Delta\tilde{r}_{i+1} = d_i$, where
\begin{align}
    a_i&= \frac{1}{4(r_i^3 - r_{i+1}^3)}\Bigg\{12 \gamma p_{i+1} r_i^2 r_{i+1}^2 + 
     M_{\star, {i+1}} \Big[ r_i^3 (2\rho_{i+1} - \rho_{i+2}) + r_{i+1}^3(\rho_{i+1} + \rho_{i+2}) - 3 r_i^2 r_{i+2} \rho_{i+1} \Big]\Bigg\}\\
    b_i&= \frac{r_i}{4(r_{i-1}^3-r_{i}^3)(r_i^3 - r_{i+1}^3)} \Bigg\{ -4 p_i [2 r_{i-1}^3 + (3\gamma -2) r_i^3](r_{i}^3 - r_{i+1}^3) \nonumber\\
    & \hspace{43mm} -4 p_{i+1} [2r_{i+1}^3 + (3\gamma -2) r_{i}^3](r_{i-1}^3 - r_i^3) \nonumber\\
    & \hspace{43mm} + 3 M_{\star,i} r_{i} \Big[r_{i+1}^3 \rho_i + r_{i-1}^3 \rho_{i+1} - r_{i}^3(\rho_i + \rho_{i+1})\Big](r_{i-1} - r_{i+1})\nonumber\\
    & \hspace{43mm} + \dfrac{d M_{\star}}{dr}\bigg|_{i} \dfrac{1}{r_{i}} (r_{i}^3 - r_{i-1}^3)(r_{i}^3 - r_{i+1}^3)(\rho_i + \rho_{i+1})(r_{i-1} - r_{i+1})\Bigg\}\\
    c_i&= \frac{1}{4(r_i^3 - r_{i+1}^3)}\Bigg\{12 \gamma p_{i+1} r_i^2 r_{i+1}^2 + 
     M_{\star, i} \Big[ r_i^3 (\rho_i + \rho_{i+1}) - 
     r_{i+1}^2 (r_{i+1} (\rho_i - 2 \rho_{i+1}) + 
        3 r_{i-1} \rho_{i+1})\Big]\Bigg\}\\
    d_i&= (p_{i+1} - p_i) r_i^2 + 
     \frac{1}{4} M_i (r_{i+1} - r_{i-1}) (\rho_i + \rho_{i+1})
\end{align}
where $\gamma=5/3$.

\subsection{Simulation Parameters}
The parameters chosen for our simulation are $\tilde{r}_{\rm min} = 3\times10^{-3}$, $\tilde{r}_{\rm max} = 5\times10^{2}$, $N_{\rm shell}=500$, and $f\sigma_m=1\,{\rm cm^2/g}$. The heat conduction time steps in the simulation are determined automatically from the system parameters at its current stage of evolution. The timestep is set by the minimum of the collision and heat-conduction timescales over all shells,  
\begin{equation}
    \delta t = \epsilon_t \min_{i=1,\cdots,N_{\rm shell}} \left\{ \frac{1}{a\rho_i \sigma_m v_i}, \; \frac{u_i}{\delta L_i / \delta M_i} \right\},
\end{equation}
where the index $i$ runs over all shells. Here, $\epsilon_t$ is a small parameter chosen to be $10^{-3}$, 
$u_i = \tfrac{3}{2}(p_i / \rho_i)$ is the specific energy of shell $i$, 
$\delta L_i = L_{i+1} - L_i$, and $\delta M_i = M_{i+1} - M_i$.

\color{black}

\section{Late-Time Evolution}
\label{app:late}
A common feature in all our gravothermal simulations is the formation of an isothermal core within which the temperature gradient of $\chi$ is suppressed, $|\partial_r T_m|\ll T_m/r$. Heuristically, this is due to the temperature-smoothing tendency of heat conduction, which is most efficient near the central region. We approximate the isothermal core as a ball of radius $r_{\rm iso}$ with a uniform temperature $T_m^{\rm iso}$. In general, this core is surrounded by an exterior cloud whose temperature profile is $T_m^{\rm ext}(r)$. Thus, the full temperature and density profiles read
\begin{align}
    T_m(r)&=T_m^{\rm iso}\Theta(r_{\rm iso}-r)+T_m^{\rm ext}(r)\Theta(r-r_{\rm iso})\\
    \rho_\chi(r)&=\rho_\chi^{\rm iso}\Theta(r_{\rm iso}-r)+\rho_\chi^{\rm ext}(r)\Theta(r-r_{\rm iso})
\end{align}
Once the full temperature profile $T_m(r)$ is set, all the other relevant quantities can be derived from it. For instance, the density profile is given by the hydrostatic equilibrium in Eq.~(1) of main text. We define $r_{\rm iso}$ to be the radius that satisfies $\partial_r T_m^{\rm ext}|_{r=r_{\rm iso}}=(T_m^\infty-T_m^{\rm iso})/r_{\rm iso}$. Integrating the gravothermal equations from $r=0$ to $r=r_{\rm iso}$ leads to the following heat equation for the isothermal core
\begin{align}
    \dot{T}_m^{\rm iso}C_{\rm iso}=L_{\rm iso}\label{eq:simpheat}
\end{align} 
The effective heat capacity of the isothermal core $C_{\rm iso}$, the luminosity emerging from the isothermal core $L_{\rm iso}$, the isothermal-core density $\rho_\chi^{\rm iso}(r)$, and isothermal-core radius $r_{\rm iso}$ are given by
\begin{align}
    C_{\rm iso}&\equiv \int_0^{r_{\rm iso}}4\pi r^2 dr\rho_\chi^{\rm iso}\left(\frac{3}{2}-\frac{\partial \ln\rho_\chi^{\rm iso}}{\partial\ln T_m^{\rm iso}}\right)=\int_0^{r_{\rm iso}}4\pi r^2 dr\rho_\chi^{\rm iso}\left(2\eta_{\rm iso}+\frac{1}{2}+\frac{-\Phi}{T_m^{\rm iso}}\right) \label{eq:Ciso}\\
    L_{\rm iso}&\equiv4\pi r_{\rm iso}^2\left[\kappa_{\rm eff}^{\rm ext}\partial_rT_m^{\rm ext}\right]_{r=r_{\rm iso}}=-\frac{6\pi C_{\rm lmfp}\sigma_m r_{\rm iso}^3\left[\rho_\chi(r_{\rm iso})\right]^2\sqrt{T_m^{\rm iso}}(T_m^{\rm iso}-T_m^\infty)}{1+(C_{\rm lmfp}/C_{\rm smfp})[\rho_\chi(r_{\rm iso})\sigma_m r_{\rm iso}]^2}\label{eq:Liso}\\
    \rho_\chi^{\rm iso}(r)&=\left(\rho_\chi^\infty\frac{T_m^\infty}{T_m^{\rm iso}}e^{-\int_\infty^{r_{\rm iso}}\frac{dr'}{r'}\frac{GM_\star/r'}{T_m(r')}}\right)e^{-\frac{\Phi(r)-\Phi(r_{\rm iso})}{T_m^{\rm iso}}}=\rho_\chi^\infty (2\eta_{\rm iso})^{2\bar{\eta}}\left(\frac{T_m^{\rm iso}}{T_m^\infty}\right)^{2\bar{\eta}-1}e^{-\frac{\Phi(r)}{T_m^{\rm iso}}-2\eta_{\rm iso}}\label{eq:rhoiso2}\\
    r_{\rm iso}&=\frac{GM_\star}{2\eta_{\rm iso}T_m^{\rm iso}}\label{eq:riso}
\end{align}
In writing the above, we defined
\begin{align}
    \eta\equiv \frac{GM_\star/r}{2T_m^{\rm ext}},\quad \eta_{\rm iso}\equiv \eta(r_{\rm iso}),\quad \bar{\eta}\equiv \frac{\int_{r_{\rm iso}}^{\infty}\eta \,dr/r}{\int_{r_{\rm iso}}^{\lambda_{\rm grav}}dr/r} \label{eq:etadefinitions},\quad \lambda_{\rm grav}&\equiv \frac{GM_\star}{T_m^\infty}
\end{align}
where have assumed $\Phi(r_{\rm iso})\approx -GM_\star/r_{\rm iso}$. In solving Eq.~\eqref{eq:simpheat}, we will choose the initial condition such that $T_m^{\rm iso}(t=0)\approx T_m^{\rm ad}(r=R_\star)$. Nevertheless, the late-time profiles are insensitive to the precise choice of initial condition.

\begin{figure}[t!]
    \centering
    \includegraphics[width=0.7\linewidth]{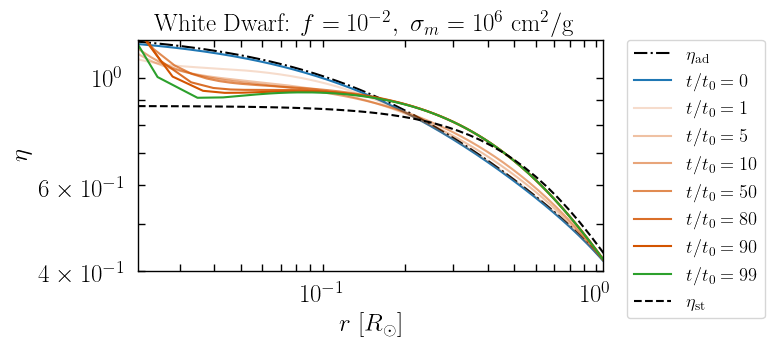}
    \caption{Snapshots of simulated $\eta$ profile (defined in Eq.~\ref{eq:etadefinitions}) for a subcomponent pile around the benchmark white dwarf of Table~\ref{tab:co_sources_sm}. The radius $r$ from the center of the white dwarf ranges from $r_{\rm iso}$ (defined in Eq.~\ref{eq:riso}) to $\lambda_{\rm grav}$ (defined in Eq.~\eqref{eq:etadefinitions}. The scale time $t_0$ is defined in Eq.~\eqref{eq:scalet}. Also shown are the adiabatic and steady profiles, $\eta_{\rm ad}(r)$ and $\eta_{\rm st}(r)$, defined in Eq.~\eqref{eq:etaad} and \eqref{eq:etast}. It can be seen that the simulated $\eta$ profile begins with $\eta_{\rm ad}(r)$ and approaches $\eta_{\rm st}(r)$.}
    \label{fig:eta}
\end{figure}

The exterior profiles, $T_m^{\rm ext}$ and $\rho_\chi^{\rm ext}$, are in principle history-dependent and can only be found by evolving the initial adiabatic profiles with the gravothermal equations. However, our simulations show that the exterior profiles evolve in a rather simple way: they interpolate between two profiles, (1) the initial \textit{adiabatic} profile and (2) a uniform-luminosity solution we call the \textit{steady} profile. The associated temperature profiles can be approximated as
\begin{align}
    T_m^{\rm ad}(r)&=T_m^\infty+\frac{2GM_\star}{5r}e^{-\frac{r^2}{\lambda_{\rm grav}^2}},\quad\phantom{ll}\text{(adiabatic)}\label{eq:Tad}\\ 
    T_m^{\rm st}(r)&=T_m^\infty+\frac{4GM_\star}{7r}e^{-\frac{7r}{4\lambda_{\rm grav}}},\quad \text{ (steady)}\label{eq:Tst}
\end{align}
At $r\gg R_\star$, these profiles correspond to
\begin{align}
    \eta_{\rm ad}(r)&=\frac{5/4}{e^{-\frac{r^2}{\lambda_{\rm grav}^2}}+\frac{5r}{2\lambda_{\rm grav}}},\quad \eta_{\rm ad}^{\rm iso}\approx\frac{5/4}{e^{-\frac{4(T_m^{\infty})^2}{25(T_m^{\rm iso}-T_m^\infty)^2}}+\frac{T_m^\infty}{T_m^{\rm iso}-T_m^\infty}},\quad \bar{\eta}_{\rm ad}\approx \frac{5}{4}\left[1+\frac{\ln\left(0.46+\frac{r_{\rm iso}}{\lambda_{\rm grav}}\right)}{\ln(\lambda_{\rm grav}/r_{\rm iso})}\right]\approx 0.9-1.1\label{eq:etaad}\\
    \eta_{\rm st}(r)&=\frac{7/8}{e^{-\frac{7r}{4\lambda_{\rm grav}}}+\frac{7r}{4\lambda_{\rm grav}}}, \quad \eta_{\rm st}^{\rm iso}\approx \frac{7/8}{e^{-\frac{T_m^\infty}{T_m^{\rm iso}-T_m^\infty}}+\frac{T_m^\infty}{T_m^{\rm iso}-T_m^\infty}},\quad \bar{\eta}_{\rm st}\approx \frac{7}{8}\left[1+\frac{\ln\left(1.15+\frac{r_{\rm iso}}{\lambda_{\rm grav}}\right)}{\ln(\lambda_{\rm grav}/r_{\rm iso})}\right]\approx0.9\label{eq:etast}
\end{align}
The expressions of $\eta_{\rm ad}$ and $\eta_{\rm iso}$ are exact for the temperature profiles in Eq.~\eqref{eq:Tad}\&\eqref{eq:Tst}; $\eta_{\rm ad}^{\rm iso}$ and $\eta_{\rm st}^{\rm iso}$ are valid in the limit $r_{\rm iso}\ll \lambda_{\rm grav}$; $\bar{\eta}_{\rm ad}$ and $\bar{\eta}_{\rm st}$ do not have closed-form expressions, so we instead show functions that provide good fits to their numerical values. We note that all these $\eta$'s are close to unity, especially when $T_m^{\rm iso}/T_m^\infty$ is not exceedingly large. This justifies the simplifying assumption we made in the main text, where we took $\eta_{\rm iso}=\bar{\eta}=\eta_{\rm eff}=1$. We display in Fig.~\ref{fig:eta} how the simulated $\eta$ profile evolves from $\eta_{\rm ad}(r)$ toward $\eta_{\rm st}(r)$.

\begin{figure}[t!]
    \centering
    \includegraphics[width=0.7\linewidth]{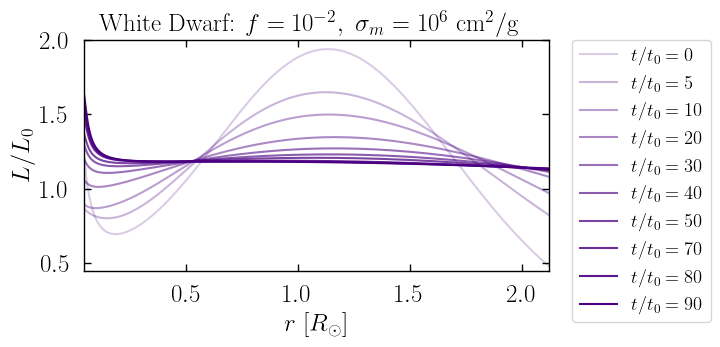} 
    \caption{Snapshots of simulated outward heat-conduction luminosity $L$ profiles (defined in Eq.~\eqref{eq:gravothermal2}) for a subcomponent pile around the benchmark white dwarf of Table~\ref{tab:co_sources_sm}. The radius $r$ from the center of the white dwarf ranges from $r_{\rm iso}$ (defined in Eq.~\ref{eq:riso}) to $\lambda_{\rm grav}$ (defined in Eq.~\eqref{eq:etadefinitions}. The scale quantities $L_0$ and $t_0$ are defined in Eqs.~\eqref{eq:scaleluminosity},\eqref{eq:scalerho}\eqref{eq:scalev},\eqref{eq:scalet}, and \eqref{eq:scalesigma}. It can be seen that the $L$ profile evolves toward and settles at a spatially-uniform steady-state.}
    \label{fig:luminosity}
\end{figure}

Let us now describe the origin of the steady profile $T_m^{\rm st}(r)$ in Eq.~\eqref{eq:Tst}, which is valid when the external profiles are in the long mean free path (LMFP) regime, $\lambda_{\rm mfp}\gg r$. As can be seen in Fig.~\ref{fig:luminosity}, the simulated luminosity $L$ evolves from the initial profile toward uniformity. The temperature profile in Eq.~\eqref{eq:Tst} corresponds to this eventual uniform-luminosity steady state. Eliminating $\rho_\chi$ from the hydrostatic equilibrium in Eq.~(1) of main text and assuming the LMFP luminosity is uniform $L\propto r^4\rho_\chi^2\sqrt{T_m^{\rm ext}}\partial_rT_m^{\rm ext}=\text{uniform}$, we find the following differential equation for the external temperature profile $T_m^{\rm ext}$
\begin{align}
    \frac{\partial_r^2T_m^{\rm ext}}{\partial_rT_m^{\rm ext}}-\frac{3\partial_r T_m^{\rm ext}}{2T_m^{\rm ext}}+\frac{4}{r}\left(1-\frac{GM_\star/r}{2T_m^{\rm ext}}\right)=0
\end{align}
The solutions to this nonlinear differential equation can be found numerically with techniques such as the shooting method, and are well fitted by Eq.~\eqref{eq:Tst}. A special case of this solution can be derived with the following simple scaling argument. At radii $r$ intermediate between $r_{\rm iso}$ and $\lambda_{\rm grav}$, we expect the solution to be scale-free. Substituting a power-law ansatz to the above equation $T_m^{\rm ext}=Cr^{-n}$, we find that $T_m^{\rm st}\approx 4GM_\star/7r$ (which implies $\rho_\chi^{\rm ext}\propto r^{-3/4}$) is the only solution. Eq.~\eqref{eq:Tst} is a generalization of this scale-free solution that captures the behavior at $r\sim r_{\rm iso}$ and $r\sim \lambda_{\rm grav}$ more accurately.

The tendency of the temperature profile to approach $T_m^{\rm ext}$ can be understood partially as follows. Initially, at $t\ll \tau_{\rm cond}(r_{\rm iso})$, heat transfer in the exterior cloud is still inefficient, and so the exterior profiles remain adiabatic, $T_m^{\rm ext}(r)=T_m^{\rm ad}(r)$ and $\rho_\chi^{\rm ext}(r)=\rho_\chi^{\rm ad}(r)$.  Then the gravothermal equations dictate that the direction of evolution is such that the magnitude of the luminosity gradient $|\partial_rL|$ decreases over time. As this happens, the rate at which the specific entropy increases, $\dot{s}\propto -\partial_r L$, becomes slower. Eventually the evolution of $s$ and hence the entire exterior profile essentially freezes when $|\partial_r L|$ is small, i.e., when $L$ is approximately uniform. In short, the uniform-luminosity solution appears to be an attractor because the system initially approaches it and slows down when close to it. Uniform-luminosity solutions are also found both analytically and numerically, as dynamical attractors, in other astrophysical systems with similar physics, including the interior of various stars. The $\rho_\chi\propto r^{-3/4}$ can be thought of as a velocity-independent $\sigma_m$ analog of the well-known Bahcall-Wolf solution \cite{Shapiro:2014oha}, which applies for a gas with a Coulombic velocity-dependent cross-section $\sigma_m\propto v^{-4}$. 

\begin{figure}[t!]
    \centering   
    \includegraphics[width=0.48\linewidth]{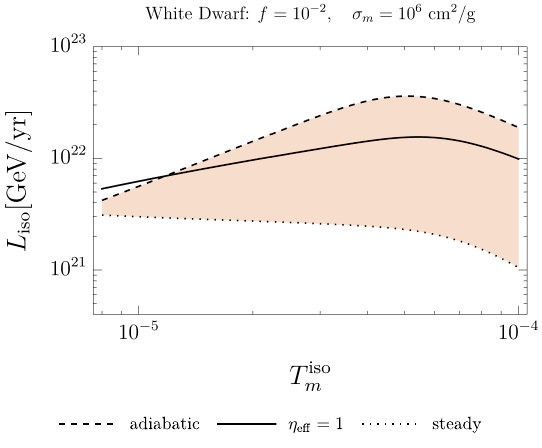} 
    \includegraphics[width=0.48\linewidth]{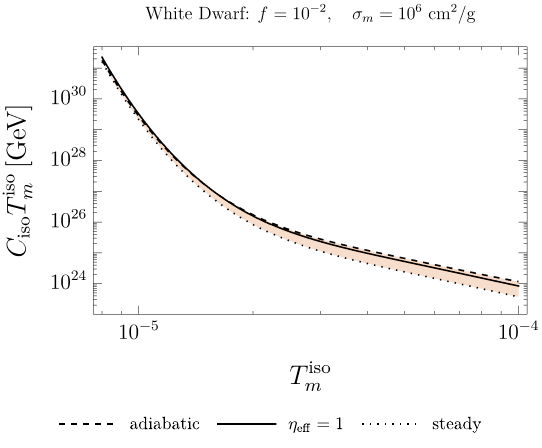} 
    \caption{Thermal properties of the isothermal core of a pile of subcomponent with $(f,\sigma_m)=(10^{-2},10^6\text{ cm}^2/\text{g})$ around the benchmark white dwarf of Table~\ref{tab:co_sources_sm}. \textit{Left:} Outward heat conduction luminosity emerging from the isothermal core $L_{\rm iso}$ (defined in Eq.~\eqref{eq:Liso}) as a function of the isothermal-core temperature $T_m^{\rm iso}$. \textit{Right:} The product between the isothermal core's heat capacity (defined in Eq.~\eqref{eq:Ciso}) and temperature $C_{\rm iso}T_m^{\rm iso}$ as a function of $T_m^{\rm iso}$. The dashed, dotted, and solid lines represent three assumptions of $\eta\equiv r\partial_r\Phi/2T_m^{\rm ext}$, namely the adiabatic and steady external profiles as defined in Eqs.~\eqref{eq:Tad}\&\eqref{eq:Tst} and $\eta_{\rm iso}=\bar{\eta}=\eta_{\rm eff}=1$ assumed in the main text, with $\eta_{\rm iso}=\bar{\eta}$ defined in Eq.~\eqref{eq:etadefinitions}.}
    \label{fig:LisoCiso}
\end{figure}

\begin{figure}[t!]
    \centering
    \includegraphics[width=0.495\linewidth]{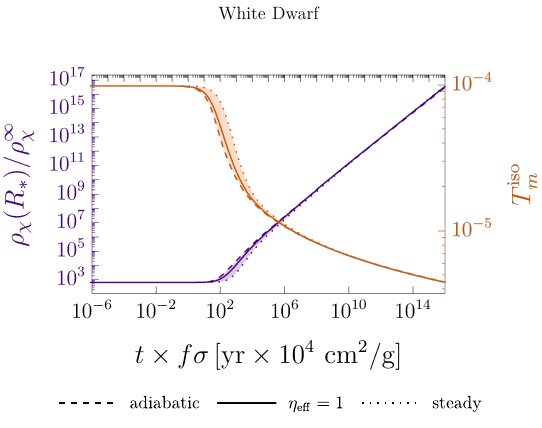}   
    \includegraphics[width=0.495\linewidth]{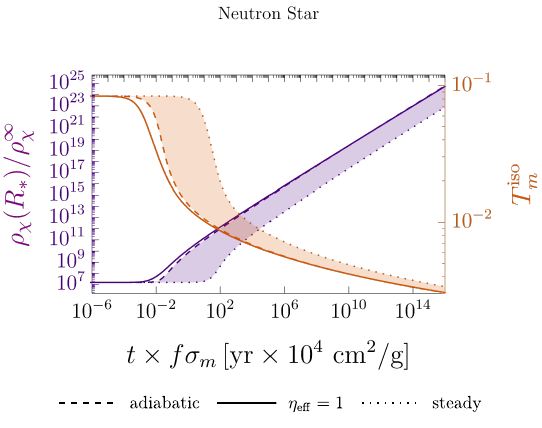} 
    \caption{Evolution of the density enhancement factor $\rho_\chi(R_*)/\rho_\chi^\infty$ of a subcomponent pile around the benchmark WD (\textit{left}) and NS (\textit{right}) of Table~\ref{tab:co_sources_sm} evaluated at the respective compact-object radii $R_\star$. The plot can be read as $\rho_\chi(R_*)/\rho_\chi^\infty$ as a function of time $t$ for a fixed $f\sigma$, e.g., fixing $f\sigma_m=10^{4}\text{ cm}^2/\text{g}$ it shows the $\rho_\chi(R_*)/\rho_\chi^\infty$ as a function of time $t$ in yr. Alternatively the plot can be read as $f^{-1}\rho_\chi(R_*)$ as a function of $f\sigma_m$ at a fixed elapsed time $t$, e.g., it could be read as showing the $\rho_\chi(R_*)/\rho_\chi^\infty$ of a 10 Gyr compact object as a function of $f\sigma_m=\# 10^{-6}\text{ cm}^2/\text{g}$, where $\#$ is the value on the $x$ axis. The results are obtained by integrating Eq.~\eqref{eq:simpheat} using the isothermal-core properties given in Eqs.~\eqref{eq:Ciso}, \eqref{eq:Liso}, \eqref{eq:rhoiso2}, \eqref{eq:etaad}, and \eqref{eq:etast}. The dashed, dotted, and solid lines represent three assumptions of $\eta\equiv r\partial_r\Phi/2T_m^{\rm ext}$, namely the adiabatic and steady external profiles as defined in Eqs.~\eqref{eq:Tad}\&\eqref{eq:Tst} and $\eta_{\rm iso}=\bar{\eta}=\eta_{\rm eff}=1$ assumed in the main text, with $\eta_{\rm iso}$ and $\bar{\eta}$ defined in Eq.~\eqref{eq:etadefinitions}.}
    \label{fig:adstevolution}
\end{figure}

In general, the external temperature profile $T_m^{\rm ext}(r)$ is somewhere between adiabatic profile $T_m^{\rm ad}(r)$ and the steady profile $T_m^{\rm st}$. We plot the corresponding heat capacity $C_{\rm iso}$ of the isothermal core and the luminosity emerging from it $L_{\rm iso}$ in Fig.~\ref{fig:LisoCiso}. 
For the most part, the $L_{\rm iso}$ is larger for the adiabatic profile than for the steady profile, while their $C_{\rm iso}$ are nearly the same. These amount to the temperature evolution rate $\dot{T}_m^{\rm iso}/T_m^{\rm iso}=-L_{\rm iso}/(C_{\rm iso}T_m^{\rm iso})$ being faster in the adiabatic case. To show the range of possible evolutions, we numerically integrate the simplified heat equation, Eq.~\eqref{eq:simpheat}, assuming these exterior solutions as two extreme cases. The resulting time-evolution of the central density $\rho_\chi(R_\star)$ is displayed in Fig.~\ref{fig:adstevolution}. In addition, we include in the same figure the central density evolution corresponding to the simplification $\eta_{\rm iso}=\bar{\eta}=\eta_{\rm eff}=1$ used in the main text. All the solutions imply that as $T_m^{\rm iso}$ cools down, the central density profile $\rho_\chi(R_\star)$ rises approximately linearly in time, increasing the heat capacity of the isothermal core $C_{\rm iso}$ exponentially, and making it progressively harder to cool $T_m^{\rm iso}$ further. The effect of $\eta^{\rm iso}$ enters the isothermal-core density $\rho_\chi^{\rm iso}$, as given in Eq.~\eqref{eq:rhoiso2}, as a mild prefactor $\eta_{\rm iso}^{\bar{\eta}}e^{-2\eta_{\rm iso}}=\mathcal{O}(1)$. The $\bar{\eta}$, on the other hand, could change $\rho_\chi^{\rm iso}$ more drastically, through the factor $(T_m^{\rm iso}/T_m^\infty)^{2\bar{\eta}-1}$. Notice that small differences in $\bar{\eta}$ is amplified when $T_m^{\rm iso}\gg T_m^\infty$. This explains why the isothermal-core density vs time evolution, shown in Fig.~\ref{fig:adstevolution}, for adiabatic and steady exterior profiles deviate more in the NS case (where $T_m^{\rm iso}\sim 10^{-3}-10^{-1}$) compared to the WD case (where $T_m^{\rm iso}\sim 10^{-6}-10^{-4}$). We note that while we expect the true evolution of $\rho_\chi(R_*)$ (as would be found in a gravothermal simulation) to lie between the dotted and dashed lines, the $\bar{\eta}=1$ line is merely a simplifying assumption, and does not have to behave in a way intermediary between the adiabatic and steady ones.

\begin{figure}[t!]
    \centering
    \includegraphics[width=0.49\linewidth]{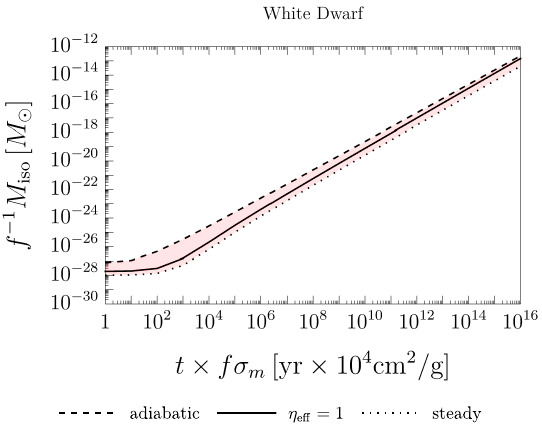}
    \includegraphics[width=0.49\linewidth]{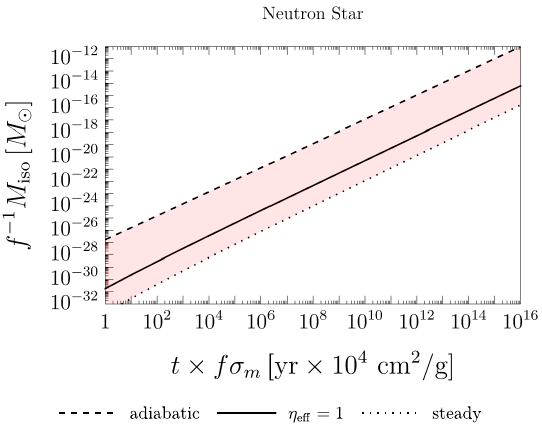}
    \caption{Evolution of the total mass of the isothermal core, $M_{\rm iso}\equiv\int_0^{\rm r_{\rm iso}} 4\pi r^2dr\, \rho_\chi^{\rm iso}$, rescaled by $f^{-1}$ around the benchmark WD (\textit{left}) and NS (\textit{right}) of Table~\ref{tab:co_sources_sm}. The plot can be read as the function $f^{-1}M_{\rm iso}(t)$ for a fixed $f\sigma$, e.g., fixing $f\sigma_m=10^{4}\text{ cm}^2/\text{g}$ it shows the $f^{-1}M_{\rm iso}$ accumulated as a function of time $t$ in yr. Alternatively the plot can be read as the function $f^{-1}M_{\rm iso}(f\sigma_m)$ for a fixed elapsed time $t$, e.g., it could be read as showing the $f^{-1}M_{\rm iso}$ accumulated around a 10 Gyr compact object as a function of $f\sigma_m=\# 10^{-6}\text{ cm}^2/\text{g}$, where $\#$ is the value on the $x$ axis. The dashed, dotted, and solid lines represent three assumptions of $\eta\equiv r\partial_r\Phi/2T_m^{\rm ext}$, namely the adiabatic and steady external profiles as defined in Eqs.~\eqref{eq:Tad}\&\eqref{eq:Tst} and $\eta_{\rm iso}=\bar{\eta}=\eta_{\rm eff}=1$ assumed in the main text, with $\eta_{\rm iso}$ and $\bar{\eta}$ defined in Eq.~\eqref{eq:etadefinitions}.}
    \label{fig:Misoevolution}
\end{figure}

In Fig.~\ref{fig:Misoevolution} we plot the total mass of the isothermal core $M_{\rm iso}\equiv\int_0^{\rm r_{\rm iso}} 4\pi r^2dr\, \rho_\chi^{\rm iso}$ as a function of time. It shows that, for the most part, $M_{\rm iso}\propto f\sigma_m t$.

\section{Pre-Gravothermal Physics}
\label{app:preGravothermal}

In our gravothermal analysis, we assume as an initial condition that the exterior temperature and density profiles are given by the adiabatic profiles, Eq.~(3). Such an initial condition is expected if we confine ourselves within the gravothermal fluid formalism. Here, we argue that similar initial conditions are expected in more general frameworks. We will simply and conservatively suppose that the potential well of the compact object of interest (WD or NS) forms suddenly, while keeping the ambient subcomponent unperturbed as this happens. In reality, the initial relaxation of the subcomponent gas toward a quasi-hydrostatic equilibrium  occurs already \textit{during} the formation of the gravitational well. Our description below applies when the collisional timescale $\tau_{\rm col}$ is much greater than the formation timescale of the compact object, which is typically short (core collapse supernova, for example, occurs in about 1 second).

First, we describe the initial relaxation in the (dynamical) fluid picture. The sudden appearance of a deep gravitational well prompts the ambient subcomponent gas to rearrange itself to establish a hydrostatic equilibrium. During this process, the bulk velocity of the infalling cold subcomponent gas accelerates and, at some point, crosses the sound speed, whereupon the gas shocks, bringing it to an abrupt halt, converting the bulk kinetic energy to random thermal energy, and increasing the entropy of the gas. The post-shock properties of the gas are given by the Rakine-Hugoniot relations \cite{2017mcpo.book.....T}, which are nothing but the continuity of mass, energy, and momentum. In a strong shock, the density jumps by a factor of 4 while the bulk kinetic energy of the flow converts entirely into thermal energy. The resulting $\rho$ and $T$ profiles are different from the adiabatic ones we assumed, but only by only $\mathcal{O}(1)$ factors.

It is instructive to also look at the initial formation in the particle picture. We start by supposing that there is no $\chi$ particle bound to the potential well. A density of $\chi$ as high as that of the adiabatic density profile, Eq.~(3), requires populating the phase-space of bound orbits, i.e. those with negative total energy per particle. This phase space can be populated via collisions. When two particles collide inside a gravitational well, one of them typically loses $\mathcal{O}(1)$ of its energy and the other gains that energy. The particle that loses energy has an $\mathcal{O}(1)$ chance of becoming bound. At the same time collisions can also deplete bound particles. Since the momentum-space volume for bound particles, $(\Delta p)^3\sim (mv_{\rm esc})^3$, is huge compared to that of free particles, $(\Delta p)^3\sim (m\sigma_v^{\rm gal})^3$, bound particles will initially increase in number. At some point, the bound$\rightarrow$free and free$\rightarrow$bound rates become equal, and a detailed balance is established.

This rough expectation can be quantified properly with the Boltzmann equation, although we will only provide a rough sketch here. Assuming the distribution function is isotropic, the Boltzmann equation at a given point $\mathbf{r}$ in space for the process $1+2\leftrightarrow 3+4$ can be simplified as \cite{Semikoz:1995rd}
\begin{align}
    \dot{f}(E_1)\sim  m^2\sigma_m\int dE_3dE_4\,F\frac{\text{min}(v_1,v_2,v_3,v_4)}{v_1}
\end{align}
where $f(E_1)$ is the distribution function of particle 1, and $E_i=mv_i^2/2+\Phi(\textbf{r})$ and $v_i$ are the total energy and velocity of particle $i$ and
\begin{align}
    F=f(E_3)f(E_4)-f(E_1)f(E_2)
\end{align}
Let us consider processes where 1 refers to a bound particle ($E_1<0$), 2  can be either free or bound, and 3 and 4 are both free. We assume the distribution function of free particles $f_{\rm free}$ is given by the ambient Galactic value, $f_{\rm free}\sim (\rho_\chi^\infty/m)/(m\sigma_v^{\rm gal})^3$. Assuming the phase-space of bound particles is initially unfilled, $f(E_1)\sim f_{\rm bound}\approx 0$, the second term in the expression of $F$ above is negligible, and so the initial rate at which the bound phase space is populated is determined entirely by the first term, $f(E_3)f(E_4)\sim f_{\rm free}^2$. The Boltzmann equation then reads
\begin{align}
    \dot{f}_{\rm bound}\sim \dot{f}(E_1)\sim m^2\sigma_m \times m(\sigma_v^{\rm gal})^2\times m(\sigma_v^{\rm gal})^2\times f_{\rm free}^2\times \mathcal{O}(1)\sim f_{\rm free}\times \underbrace{(\rho_\chi^\infty \sigma_m \sigma_v^{\rm gal})}_{\tau_{\rm col}^{-1}}
\end{align}
where we assumed $v_1\sim v_2\sim v_3\sim v_4$ and substituted $\int dE_{3,4}\rightarrow \Delta E_{3,4}\sim m(\sigma_v^{\rm gal})^2$. Detailed balance is achieved when $F=0$, i.e. when $f_{\rm bound}= f_{\rm free}$ (analogous to the steady state of DM bound to the solar system \cite{2004PhRvD..69l3505L,1991ApJ...368..610G}). Based on the above estimate, reaching this detailed balance state takes about
\begin{align}
    \tau_{\rm col}\sim 1\text{ Gyr}\left(\frac{f\sigma_m}{1\text{ cm}^2/\text{g}}\right)^{-1}
\end{align}
Thus, the gravothermal analysis is applicable if $\tau_{\rm col}\ll t_{\rm age}=10\text{ Gyr}$, which we have already assumed in our main analysis. Once detailed balance is reached, the density of bound particles is given by
\begin{align}
    \rho_{\rm bound}\sim mf_{\rm bound}\times [m\sigma_v^{\rm bound}(r)]^3\sim m^4f_{\rm free}\times \left[-\Phi_*(r)\right]^{3/2}\sim \rho_\chi^\infty\left[\frac{-\Phi_*(r)}{T_m^\infty}\right]^{3/2}
\end{align}
where we have used the virial theorem to write the velocity dispersion of bound particles as $[\sigma_v^{\rm bound}(r)]^2\sim-\Phi_*(r)$. This reproduces the order of magnitude of the adiabatic profile assumed in the main text.

\color{black}

\section{Particle Physics Models}
\label{app:model}
Here, we present concrete particle physics models as existence proofs and discuss their relevant model-dependent phenomenology. 

\subsection{Dark Scalar}
\begin{figure}[t!]
    \centering
    \includegraphics[width=0.49\linewidth]{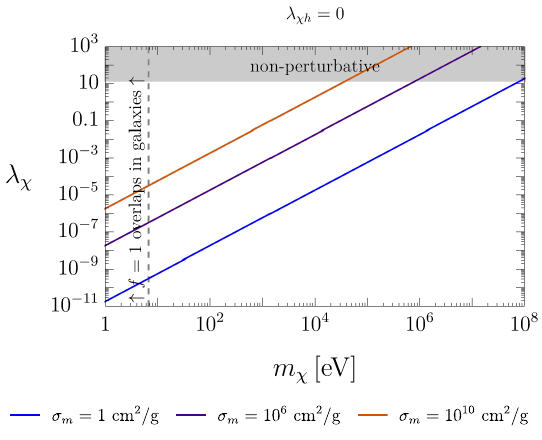}
    \includegraphics[width=0.5\linewidth]{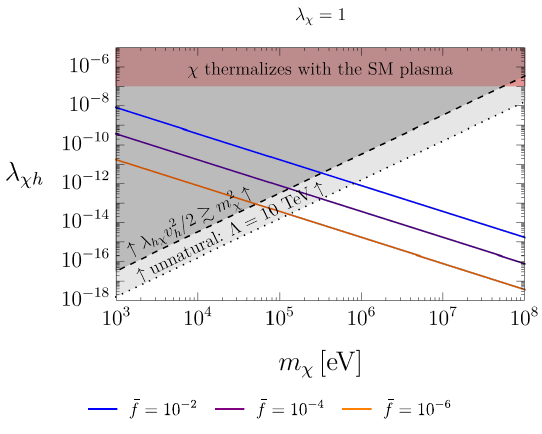}
    \caption{Dark scalar parameter space. \textit{Left:} Self-coupling $\lambda_\chi$ vs mass $m_\chi$ space in the absence of a higgs-portal, $\lambda_{\chi h}=0$. In the gray region, the $\lambda_\chi$ is non-pertubative. To the left of the dashed line the galactic population of $f=1$ scalars have occupation number greater than unity. Colored lines correspond to different values of $\sigma_m$. \textit{Right:} Higgs-portal coupling $\lambda_{\chi h}$ vs mass $m_\chi$ for self-coupling $\lambda_\chi=1$. In the red region, $\chi$ particles achieve full thermalization with the SM plasma in the early universe. In the gray regions, a tuning of parameters is needed to keep $m_\chi$ light. Color lines correspond to different cosmic DM fractions $\bar{f}$ of $\chi$ today.}
    \label{fig:scalarparameterspace}
\end{figure}

The subcomponent $\chi$ considered in the main text could be a scalar $\chi$ with the following Lagrangian
\begin{align}
    \mathcal{L}_\chi=\frac{1}{2}\left(\partial\chi\right)^2-\frac{1}{2}m_\chi^2\chi^2- \frac{\lambda_{\chi}}{4}\chi^4
\end{align}
The non-relativistic self-scattering cross section per unit mass in this model is
\begin{align}
    \sigma_m=\frac{9\lambda_{\chi}^2}{32\pi m_\chi^3}=2\times 10^{4}\lambda_{\chi}^2\text{ cm}^2/\text{g } \left(\frac{m_\chi}{\MeV}\right)^{-3}
\end{align}
To ensure vacuum stability and perturbativity, the self coupling must lie in the range $0<\lambda_{\chi}< 16\pi$, which implies that in order for the scalar to have a large $\sigma_m$ it needs to be sufficiently light. Light particles tend to have large de Broglie wavelengths and are thus prone to overlap. The phase-space density $\mathcal{F}$ of $\chi$ at galactic scales is
\begin{align}
   \mathcal{F}_{\rm gal}= \frac{\rho_\chi^\infty/m_\chi}{(m_\chi \sqrt{T_m})^{3}}\sim 0.2 f\left(\frac{T_m}{10^{-6}}\right)^{-3/2}\left(\frac{m_\chi}{10\eV}\right)^{-4}
\end{align}
During the gravothermal evolution, $\mathcal{F}$ may increase by orders of magnitude. While number-changing processes such as $\chi\chi\chi\chi\rightarrow\chi\chi$ may prevent it from reaching unity, such processes could also be turned off by making the scalar a complex field and populating it asymmetrically, with only particles and no antiparticles \cite{Petraki:2013wwa,Kaplan:2009ag}.  The scalar $\chi$ could also be composite. It could represent, e.g., dark glueballs of a confined Yang-Mills sector \cite{Boddy:2014yra,Kribs:2016cew,Jo:2020ggs} whose cubic self-interaction $\phi^3$ allows for $\chi\chi\chi\rightarrow \chi\chi$ number-changing processes. We show the parameter space of the dark scalar model in Fig.~\ref{fig:scalarparameterspace}.

If the $\chi$ particles are produced relativistically, they can subsequently self-thermalize through $2\rightarrow 4$ number-changing processes of the form $\chi\chi\rightarrow\chi\chi\chi\chi$ \cite{Bernal:2017kxu}. Assuming the $\chi$ sector is in a kinetic equilibrium with a temperature $T'$, the cross-section for this process when $\chi$ is non-relativistic and relativistic are  $\left<\sigma_{2\rightarrow 4}v\right>\sim \lambda_\chi^4/m_\chi^2$  and $\left<\sigma_{2\rightarrow 4}v\right>\propto T^{\prime -2}$, respectively, where $T'$ is the dark sector temperature. This implies that the thermalization efficiency $n_\chi\left<\sigma_{2\rightarrow 4} v\right>H^{-1}$ is maximized at $T'\sim m_\chi$. Requiring $\left.n_\chi\left<\sigma_{2\rightarrow 4} v\right>H^{-1}\right|_{T'=m_\chi}\gtrsim 1$, we find that the $\chi$ sector would self-thermalize if
\begin{align}
    \lambda_\chi\gtrsim \xi_s^{-1/6}\left(\frac{m_\chi}{M_{\rm pl}}\right)^{1/4}=5\times 10^{-6}\xi_s^{-1/6}\left(\frac{m_\chi}{\MeV}\right)^{1/4}
\end{align}
where $M_{\rm pl}=2\times 10^{18}\GeV$ is the reduced Planck mass and $\xi_s\equiv s'/s$ is the ratio between the entropy densities of $\chi$ and the SM after the $\chi$ sector is self-thermalized. This is a conserved quantity as long as the $\chi$ sector remains self-thermalized and decoupled from the SM sector. Since we are mainly interested in $\chi$ with a large self coupling, we will assume that the above self-thermalization condition is satisfied. Once thermalized, the $\chi$ sector will remain thermalized until the rate of $4\rightarrow 2$ processes $n_\chi^3\left<\sigma_{4\rightarrow 2}v^3\right>$ drops below Hubble, whereupon $n_\chi$ freezes out, typically at $T^\prime \sim m_\chi/10$. The resulting cosmic relic DM fraction of $\chi$ today is \cite{Heikinheimo:2016yds}
\begin{align}
    \bar{f}\approx 2\times 10^5 \xi_s\left(\frac{m_\chi}{\MeV}\right) \label{eq:scalarrelicf}
\end{align}
The value of $\xi_s$ depends on the way in which the $\chi$ sector is populated. Even if the $\chi$ particles are decoupled from the SM, an abundance of $\chi$ could also arise, for instance, through asymmetric reheating \cite{Berezhiani:1995am,Adshead:2016xxj}. They could also be produced through a gravitational production mechanism during or at the end of inflation \cite{Tenkanen:2019aij, Peebles:1999fz,Kolb:2023ydq}. While some of these scenarios are subject to isocurvature-perturbation constraints from CMB observations, they are much less severe for a subcomponent.

Next, we mention several consequences of coupling $\chi$ with the SM. As an example, we consider the following Higgs-portal coupling
\begin{align}
    V_{\chi h}=\frac{\lambda_{\chi h}}{2}\chi^2 \mathcal{H}^\dagger \mathcal{H}
\end{align}
where $\mathcal{H}$ is the Standard Model Higgs doublet, and we have removed the cubic Higgs-portal coupling $A\chi\mathcal{H}^\dagger \mathcal{H}$ by imposing a $\mathds{Z}_2$ symmetry, under which $\chi\rightarrow -\chi$. This implies that indirect-detection signals will be dominated by annihilation instead of decay. A consequence of introducing the above coupling is that the $\chi$ particles will receive mass-squared contributions through it. At tree level, we have $m_\chi^2=\mu_\chi^2+\lambda_{\chi h}v_h^2/2$
with $v_h=246\GeV$. We assume that $m_\chi^2>0$ in order to keep $\left<\chi\right>=0$, ensuring the stability of the $\chi$ particles. Furthermore, quantum corrections are expected to contribute $\delta m_\chi^2\gtrsim \lambda_{\chi h}\Lambda^2/16\pi^2$ to $\chi$'s mass squared. To avoid fine-tuning, we require $m_\chi^2>\text{max}\left(\lambda_{\chi h}v_h^2/2,\lambda_{\chi h}\Lambda^2/16\pi^2\right)$. A large $\sigma_m$ requires a small $m_\chi$ since $\lambda_\chi$ is limited by perturbativity. A small $m_\chi$, in turn, requires a correspondingly small $\lambda_{\chi h}$ in order to keep $m_\chi$ small. See Fig.~\ref{fig:scalarparameterspace}.

The dominant production channel of $\chi$ in the early universe is through higgs boson decay $h\rightarrow \chi\chi$, whose rate is $\Gamma_{h\rightarrow \chi\chi}=\lambda_{\chi h}^2v_{ h}^2/32\pi m_h$, where the higgs mass is $m_h=125\GeV$. These decays occur mostly at $T\sim m_h/3$. If $\Gamma_{h\rightarrow\chi\chi}H^{-1}\gtrsim 1$ at that time, which amounts to $\lambda_{\chi h}\gtrsim 10^{-7}$, then $\chi$ reaches a thermal equilibrium with the SM plasma. The later case is ruled out by the combination of collider and direct detection constraints \cite{Lebedev:2021xey,Cline:2013gha}, so we focus on the freeze-in scenario with $\lambda_{\chi h}\lesssim 10^{-7}$ here. The initial freeze-in energy density of $\chi$ can be estimated as $\rho_{h\rightarrow \chi\chi}\sim [m_h(m_h T)^{3/2}e^{-m_h/T}\Gamma_{h\rightarrow \chi\chi}H^{-1}]_{T=m_h/3}$. This leads to the following entropy ratio after self-thermalization of $\chi$
\begin{align}
    \xi_s=\frac{s'}{s}\sim \left(\frac{\rho_{h\rightarrow\chi\chi}}{\rho_{\rm crit}}\right)^{3/4}_{T=m_h/3} \sim 5\times 10^{-6}\left(\frac{\lambda_{\chi h}}{10^{-10}}\right)^{3/2}
\end{align}
From this and Eq.~\eqref{eq:scalarrelicf}, we estimate the present-epoch cosmic DM fraction of $\chi$ to be 
\begin{align}
    \bar{f}\sim \left(\frac{\lambda_{\chi h}}{10^{-10}}\right)^{3/2}\left(\frac{m_\chi}{\MeV}\right)
\end{align}

\subsection{Dark Atom}
Consider a QED-like theory with the following Lagrangian
\begin{align}
    \mathcal{L}_{\rm dQED}=-\frac{1}{4}F_{\mu\nu}^\prime F^{'\mu\nu}+
    \bar{p}^\prime\left(i\slashed{\partial}+g'\slashed{A}+m_{p'}\right)p'+\bar{e}^\prime\left(i\slashed{\partial}-g'\slashed{A}-m_{e'}\right) e'
\end{align}
This theory includes a hidden $U(1)'$ gauge field $A_\mu^\prime$ whose field strength tensor is $F_{\mu\nu}^\prime$, a dark proton $p'$ of mass $m_{p'}$ and $U(1)'$ charge $+g'$, and a dark electron $e'$ of mass $m_{e'}$ and $U(1)'$ charge $-g'$, \cite{Kaplan:2009de,Cline:2013pca,Cline:2013zca}. A dark proton and a dark electron can bind to form a hydrogen-like dark atom $\text{H}'$ with Bohr radius $a_0^\prime=(\alpha'\mu_{\rm H'})^{-1}$ and binding energy $B'=\alpha^{\prime 2} \mu_{\rm H'}/2$, where $\mu_{\rm H'}=m_{e'}m_{p'}/(m_{e'}+m_{p'})$ is the $p'$-$e'$ reduced mass and $\alpha'=g^{\prime 2}/4\pi$ is the dark analog of the fine-structure constant. If the binding energy $B'$ is considerably larger than the highest temperature of the gravothermal analysis and the ionized fraction is tiny, dark atoms $\text{H}'$ can play the role of the subcomponent with elastic self-interactions $\chi$. Two dark atoms collide elastically with 
\begin{align}
    \sigma_m\sim \frac{100a_0^2}{m_p'}=2\times 10^6\alpha'^2\text{cm}^2/\text{g}\left(\frac{m_{p'}}{\GeV}\right)^{-1}\left(\frac{m_{e'}}{100\keV}\right)^{-2}
\end{align}
for low momentum transfers and $m_{p'}\gg m_{e'}$. Notably, the composite nature of $\text{H}'$ allows it to have a large $\sigma_m$ with a relatively heavy $m_\chi= m_{H'}\approx m_{p'}$ compared to, e.g., the dark scalar model of the previous subsection. Alternatively, if the dominant DM is in the form of dark atoms, unpaired dark protons and dark electrons could play the role of subcomponents with large cross-sections, since dark-charged particles can have cross sections significantly larger than that of neutral dark atoms. The simplest way to populate the dark QED sector is, again, through asymmetric reheating or a gravitational production mechanism. CMB observations are sensitive to a dark QED sector that comprises a fraction as small as $5\%$ of the total dark matter, if it is highly ionized and undergoes dark acoustic oscillations. However, if the dark QED sector is sufficiently cold and have negligible ionization fraction, such limits are considerably weakened. In that case, the subdominant dark electrons and dark protons would have velocity-dependent cross sections. Finally, to get non-gravitational signals, one could turn on kinetic mixing $\epsilon F^{\mu\nu}F_{\mu\nu}^\prime/2$.

\bibliography{references}

\end{document}